\documentclass[range]{ar2e}
\usepackage{amsmath}
\usepackage{epsfig}
\usepackage{gribov}
\newcommand{\beq}{\begin{equation}}
\newcommand{\eeq}{\end{equation}}
\newcommand{\bea}{\begin{eqnarray}}
\newcommand{\eea}{\end{eqnarray}}
\newcommand{\non}{\nonumber}
\newcommand{\oh}{{\textstyle{\frac{1}{2}}}}
\newcommand{\J}{{\cal J}}
\newcommand{\bx}{{\mathbf{x}}}
\newcommand{\by}{{\mathbf{y}}}
\renewcommand{\d}{\delta}

\newdimen\picraise
\newcommand\picbox[1]
{
  \setbox0=\hbox{\input{#1}}
  \picraise=-0.5\ht0
  \advance\picraise by 0.5\dp0
  \advance\picraise by 3pt      
  \hbox{\raise\picraise \box0}
}

\def\LUV{\Lambda_{\mbox{\scriptsize UV}}}

\def\Re{\mathop{\rm Re}}
\def\Im{\mathop{\rm Im}}
\newcommand{\naive}{na\"{\i}ve}

\def\lrang#1{\left\langle#1\right\rangle}
\def\abs#1{\left|#1\right|}
\def\crit{\mbox{\scriptsize crit}}
\def\acr{\alpha_{\crit}}
\def\GeV{\mathop{\rm Ge\!V}}
\def\MeV{\mathop{\rm Me\!V}}
\def\cO#1{{\cal O}\left(#1\right)}
\def\Tr{\mathop{\rm Tr}}

\newcounter{hran}
\renewcommand{\thehran}{\arabic{hran}}

\def\bmini{\setcounter{hran}{\value{equation}}
\refstepcounter{hran}\setcounter{equation}{0}
\renewcommand{\theequation}{\thehran\alph{equation}}
\begin{eqnarray}}

\def\bminiG#1{\setcounter{hran}{\value{equation}}
\refstepcounter{hran}\setcounter{equation}{-1}
\renewcommand{\theequation}{\thehran\alph{equation}}
\refstepcounter{equation}\label{#1}\begin{eqnarray}}

\def\emini{\end{eqnarray}\relax\setcounter{equation}{\value{hran}}\renewcommand{\theequation}{\arabic{equation}}}

\begin{document}

\title{The Gribov Conception of Quantum Chromodynamics}
 \author{Yuri L. Dokshitzer \affiliation{LPTHE, University Paris-VI, 4
 place Jussieu, F-75252 Paris, France\\ and\\PNPI, 188350 Gatchina,
 St. Petersburg, Russia}
 Dmitri E.\ Kharzeev \affiliation{
 Physics Department, Brookhaven
 National Laboratory, 
%
 Upton,\\ New York 11973-5000, USA}}

\markboth{Dokshitzer \&\  Kharzeev}{Gribov Conception of Quantum Chromodynamics}

\begin{abstract}
  A major contribution to the quest of constructing quantum dynamics
  of non-Abelian fieds is due to V.N.~Gribov.
  Perturbative approach to the colour confinement, both in
  gluodynamics and the real world, was long considered heretic but is
  gaining ground.  We discuss Gribov's approach to the confinement
  problem, centered around the r\^ole played by light quarks --- the
  supercritical light quark confinement scenario. We also review some
  recent developments that are motivated, directly or indirectly, by
  his ideas.
\end{abstract}

\maketitle

\section{INTRODUCTION}

  One of the most challenging problems for modern theoretical physics
  is understanding the confinement of colour --- the selection rule
  suggested by quantum chromodynamics (QCD) to explain the fact that
  quarks (and gluons) appear in the physical spectrum only {\em
  imprisoned}\/ inside composite states --- colourless mesons and
  baryons.

  An amazing success of the relativistic theory of electron and photon
  fields --- quantum electrodynamics (QED) --- has produced a
  long-lasting negative impact: it taught the generations of
  physicists that came into the business in/after the 70's to ``not to
  worry''.  Indeed, today one takes a lot of things for granted.

  One rarely questions whether the alternative roads --- secondary
  quantization, functional integral and the Feynman diagram approach
  --- really lead to the same quantum theory of interacting fields.

  One feels ashamed to doubt an elegant and powerful, but potentially
  deceiving, technology of translating the dynamics of quantum fields
  into that of statistical systems (imaginary time $=$ Euclidean
  rotation trick, resulting in the {\em coupling}\/ $\leftrightarrow$
  {\em temperature}\/ analogy).

  One takes the original concept of the ``Dirac sea'' --- the picture
  of the fermionic content of the vacuum --- as an anachronistic
  model, a sort of Maxwell's mechanical ether, {\em
  I'd-rather-do-without}.

  One was taught to consider the problems with field-theoretical
  description of point-like objects and their interactions at very
  small distances (ultraviolet divergences) as purely technical: {\em
  renormalize it and forget it}.

  So far so good for QED, where the physical objects --- electrons and
  photons --- are direct images of the fundamental fields that one put
  into the local Lagrangian of the theory. In other words, {\em
  interacting}\/ fields closely resemble their {\em bare}\/
  counterparts. In these circumstances, the ground state of the theory
  --- the physical vacuum --- turns out to be, in a certain sense,
  trivial. The r\^ole played by the vacuum fluctuations of
  electromagnetic fields and by the ``Dirac sea'' of negative-energy
  electron states, is as important as it is
  straightforward. Interaction with the vacuum makes the effective
  interaction strength $\alpha_{\mbox{\scriptsize e.m.}}$ (as well as
  the effective electron mass) {\em run}\/ with the distance. At the
  same time, it does not affect the {\em nature}\/ of the interacting
  fields.
  

  Not so clear in QCD. Here the physical hadronic states are not in
  one-to-one correspondence with the fundamental quarks and gluons:
  interaction with the QCD vacuum changes the bare fields beyond
  recognition. In such an environment, our preconceived ideas about
  QFT dynamics may turn out to be a handicap, a serious hidden
  obstacle on the way to attacking the confinement problem.

  Vladimir Naumovich Gribov (1930--1997) belonged to a generation of
  physicists (now almost extinct) that {\em thought}\/ about the
  Quantum Field Theory (QFT), that was used to {\em questioning}\/ its
  foundations, concepts and means.
  An invitation to the ``Gribov conception of QCD'' is an invitation
  to {\em unlearn}. Learning to unlearn isn't easy. This possibly
  explains why the programme that Gribov formulated and was pursuing
  in the 90's of explaining the confinement of colour as
  ``supercritical binding'' of light quarks has yet to receive the
  attention it deserves from the physics community at large.

  It goes without saying that Gribov could be wrong in his vision
  about the nature of the QCD confinement. However, it does not seem
  to us a good policy to be simply indifferent to what one of the
  creators of the modern theoretical physics had to say on the
  subject, to the ideas and tools he has developed during the last
  half of his professional life.

  This review does not attempt to cover Gribov's impact on theoretical
  physics.  Suffices it to mention that his name is attached to many a
  key notion of the theory arsenal: Gribov--Froissart projection and
  the Gribov vacuum pole (Pomeron), Gribov factorization, Reggeon
  Calculus and Reggeon Field Theory (RFT), Gribov diffusion, the AGK
  (Abramovsky--Gribov--Kancheli) cutting rules, the Gribov
  bremsstrahlung theorem, Gribov--Glauber theory of relativistic
  multiple scattering, Gribov--Lipatov evolution equations, Gribov
  copies and the horizon, etc.

  Let us remark that Gribov's impact on modern physics is deeper than
  it is generally known to be. A couple of examples will illustrate
  the point.

 Working on the problem of the so-called strong-coupling regime of
 interacting Pomerons, V.~Gribov and A.~Migdal developed
  an ingenious technique for ana\-lysing dynamical systems with
  long-range fluctuations. Such fluctuations being typical for
  condensed matter physics near the critical temperature, this
  triggered an important breakthrough in solid state physics.  The
  subsequent works of A.~Migdal and A.~Polyakov, and a contemporary
  more general treatment suggested by L.~Kadanoff and K.~Wilson, have
  established the scaling solution of the problem of the second order
  phase transitions.

 In 1969, in one of his jewels {\em Interaction of photons and
 electrons with nuclei at high energies}\/
 Gribov established and described the space-time picture of particle
 interactions at high energies.
 This work found a way through the {\em iron curtain}.\footnote{It was
 reported, prior to publication, by J.~Bjorken at Caltech seminar.}
 Its key elements were incorporated into the famous Feynman book which
 laid the foundation of the parton model. The Feynman--Gribov parton
 model, that is.

 The last 20 years of his life V.~Gribov devoted to non-Abelian
 quantum gauge theories, and to the QCD confinement problem in
 particular.
 His very first QCD study of 1976-77 produced a brilliant physical
 explanation of asymptotic freedom introduced into the physics of
 strong interactions by D.\ Gross, F.\ Wilczek and H.\ Politzer.
 Gribov's approach was based on an early observation of the
 anti-screening phenomenon made by I.~Khriplovich in a pre-historic
 1969, and revealed the inconsistency of the standard
 field-theoretical treatment of gluon fields (Gribov copies).
  In the late 80's Gribov suggested the quark confinement scenario
  based on the so-called supercritical binding of light fermions by a
  quasi-Coulomb interaction.

  His last works remain to be discovered, understood and developed.

 For V.N.~Gribov, the problem of confinement as the problem of
 understanding the dynamics of vacuum fluctuations, and of the
 structure of hadrons as physical states of the theory, was always
 inseparable from the problem of understanding the physics of high
 energy hadron scattering (the Pomeron picture).

  Therefore, we start by an overview of the Gribov's development of
  high-energy physics. The two subsequent Sections are devoted to his
  contribution to the quantum dynamics of non-Abelian Yang--Mills
  gauge fields in general, and to the basics of the Gribov light-quark
  confinement picture. The last Section briefly describes the {\em
  byproducts}\/ of the QCD programme: an unpublished study of QED at
  short distances (a potential resolution of the notorious ``Landau
  ghost'' problem) and his picture of a composite (super-critically
  bound) Higgs boson.

\section{HADRON INTERACTIONS AT HIGH ENERGIES}

 In the late fifties, when Vladimir Gribov, then a young researcher at
 the Ioffe Physico-Technical Institute (Leningrad, USSR) became
 interested in the physics of strong hadron interactions, not much was
 understood about processes at high energies.  The only theoretical
 result, derived from the first principles, was the Pomeranchuk
 theorem --- an asymptotic equality of particle and antiparticle total
 scattering cross sections.

 \subsection{Asymptotic Behaviour $s^{\alpha(t)}$}

 Gribov's 1960 paper {\em Asymptotic behaviour of the scattering
 amplitude at high energies}~\cite{1960} was a breakthrough. Using the
 so-called double-dispersion representation for the scattering
 amplitude, suggested by S.\ Mandelstam back in 1958~\cite{Mand},
 Gribov proved an inconsistency of the then popular {\em black disk
 model}\/ of hadron-hadron scattering. This analysis may be considered
 as the first building block put into the edifice of the modern theory
 of hadron interactions. It has demonstrated the combined power of the
 general principles of relativistic quantum theory --- unitarity
 (conservation of probability), analyticity (causality) and the
 relativistic nature (crossing symmetry) --- as applied to high-energy
 scattering phenomena.

 By studying the analytic properties in the cross channels, he showed
 that the imaginary part of the scattering amplitude in the form
\begin{equation} \label{blackdisk}
   A_1(s,t) = s\,f(t)
\end{equation}
 that constituted the black disk model of diffraction in the physical
 region of $s$-channel scattering, contradicts the unitarity relation
 for partial waves in the crossing $t$-channel. To solve the puzzle,
 Gribov suggested the behaviour of the amplitude (for large $s$ and
 finite $t$) in the general form
\begin{equation}
\label{Fst}
   A_1(s,t)= s^{q(t)}\,B_t(\ln s) \,,
\end{equation}
 where $B_t$ is a slow (non-exponential) function of $\ln s$
 (decreasing fast with $t$) and $q(0)=1$ ensures the approximate
 constancy of the total cross section with energy.

 In this first paper Gribov only analysed the constant exponent,
 $q(t)=1$, having remarked on the possibility $q(t)\neq\mbox{const}$
 as ``{\em extremely unlikely}''. Indeed, considering the
 $t$-dependence of the scattering amplitude, this would correspond to
 a strange picture of the radius of a hadron infinitely increasing
 with energy.\footnote{Reportedly, it was L.D.~Landau who ``forced''
 Gribov to publish Eq.~\eqref{Fst} in its general form, with
 $q(t)\neq\mbox{const}$. ``{\em You are too young to judge}\/'' the
 blessing went, according to legend.}
 (In this particular case he proved that the cross section has to
 decrease at high energies, $B_t(\ln s)< 1/\ln s$, to be consistent
 with the $t$-channel unitarity.)

 Soon after that Gribov became aware of the finding by
 T.~Regge~\cite{Regge} that in {\em non-relativistic quantum
 mechanics}, in the unphysical region $|t|\gg s$ (momentum transfer
 much larger than the energy, corresponding to large {\em imaginary}\/
 scattering angles $|\cos\Theta|\to\infty$), the scattering amplitude
 has a form
\begin{equation}\label{Reggeamp}
    A(s,t) \>\propto\> t^{\ell(s)}\,,
\end{equation}
 where $\ell(s)$ is the {\em position of the pole}\/ of the partial
 wave $f_\ell$ in the complex plane of the orbital momentum
 $\ell$.\footnote{ Gribov apparently learned about the Regge result
 from a paper by G.~Chew and S.~Frautschi
 of 1960 which contained a {\em footnote}\/ describing the main Regge
 findings.}

 T.~Regge found that the poles of the amplitude in the complex
 $\ell$-plane were intimately related with bound states/resonances.
 It is this aspect of the Regge behaviour that initially attracted the
 most attention:
\begin{quote}
 ``{\em S.\ Mandelstam has suggested and emphasized repeatedly since
 1960 that the Regge behaviour would permit a simple description of
 dynamical states (private discussions). Similar remarks have been
 made by R.~Blankenbecker and M.L.~Goldberger and by K.~Wilson.}''~\cite{quot}
\end{quote}
 The structure of the Regge amplitude Eq.~\eqref{Reggeamp} motivated
 Gribov to return to the consideration of the case of $t$-dependent
 exponent in his general high-energy ansatz Eq.~\eqref{Fst} that was
 dictated by $t$-channel unitarity.
 His letter to ZhETF {\em Possible asymptotic behaviour of elastic
 scattering}~\cite{letter} became the first application of Regge ideas
 to the high-energy asymptotic behaviour of scattering amplitudes.

 By then M.~Froissart had already proved his famous theorem that
 limits the asymptotic behaviour of the total cross
 sections~\cite{Froissart},
\begin{equation}\label{stot}
 \sigma^{\mbox{\scriptsize tot}} \propto s^{-1}\left| A_1(s,0)\right|
 \><\>  C\, \ln^2 s\,.
\end{equation}
 Thus, having accepted $\ell(0)=1$ for the rightmost pole in the
 $\ell$-plane --- the {\em vacuum pole}\/ --- as the condition ``{\em
 that the strongest possible interaction is realized}\/'', Gribov
 formulated and discussed
 ``{\em the main properties of such an asymptotic scattering behaviour
 \ldots which in spite of having a few unusual features is
 theoretically feasible and does not contradict the experimental
 data}\/'':
\begin{itemize}
\item
  the total interaction cross section is constant at high energies,
\item
  the elastic cross section tends to zero as $1/\ln s$,
\item
  the scattering amplitude is essentially imaginary,
\item
 the significant region of momentum transfer in elastic scattering
 shrinks with energy increasing, $\sqrt{-t}\propto (\ln s)^{-1/2}$.
\end{itemize}
 He also analysed the $s$-channel partial waves to show that for small
 impact parameters $\rho<R$ their amplitudes fall as $1/\ln s$, while
 the interaction radius $R$ increases with energy as
 $\rho\propto\sqrt{\ln s}$.  He concluded:
\begin{quote}
 ``{\em this behaviour means that the particles become grey with
 respect to high-energy interaction, but increase in size, so that the
 total cross section remains constant}.''
\end{quote}
 Thus, {\em shrinkage of the diffractive peak}\/ was predicted, which
 was then experimentally verified at particle accelerator experiments
 in Russia (IHEP, Serpukhov), Switzerland (CERN) and the US (FNAL,
 Chicago), as were the general relations between the cross sections of
 different processes, that followed from the {\em Gribov factorization
 theorem}~\cite{GP}.

 These were the key features of what has become known, quite
 ironically, as the ``Regge theory'' of strong interactions at high
 energies.

 On the opposite side of the {\em 
 curtain}, the basic properties of the Regge pole picture of
 forward/backward scattering were formulated half a year later by
 G.~Chew and S.~Frautschi~\cite{CF}.  In particular, they suggested
 ``{\em the possibility that the recently discovered $\rho$ meson is
 associated with a Regge pole whose internal quantum numbers are those
 of an $I=1$ two-pion configuration},'' and conjectured the universal
 high-energy behaviour of backward $\pi^+\pi^0$, $K^+K^0$ and $pn$
 scattering due to $\rho$-reggeon exchange.  Chew and Frautschi also
 stressed that the hypothetical Regge pole with $\alpha(0)=1$
 responsible for forward scattering possesses quantum numbers of the
 {\em vacuum}.

 Dominance of the Gribov vacuum pole automatically validates the
 Pomeranchuk theorem.  The name ``Pomeron'' for the vacuum pole was
 coined by Murray Gell-Mann, who referred to Geoffrey Chew as an
 inventor.

\subsection{Complex Angular Momenta in Relativistic Theory}

 In non-relativistic quantum mechanics the interaction Hamiltonian
 allows for scattering partial waves to be considered as analytic
 functions of complex angular momentum $\ell$.
%
 Gribov's paper {\em Partial waves with complex orbital angular
 momenta and the asymptotic behaviour of the scattering amplitude}\/
%
%
 showed that the partial waves with complex angular momenta can be
 introduced in a relativistic theory as well.
%
 Here it is the {\em unitarity}\/ in the crossing channel that {\em
 replaces the Hamiltonian}\/ and leads to analyticity of the partial
 waves in $\ell$.  The corresponding construction is known as the
 ``Gribov--Froissart projection''~\cite{GF}.

 Few months later
%
%
 Gribov demonstrated that the simplest (two-particle) $t$-channel
 unitarity condition indeed generates {\em moving  poles}\/
 in the complex $\ell$-plane.  This was the {\em proof}\/ of the Regge
 hypothesis in relativistic theory~\cite{poles}.

 The ``Regge trajectories'' $\alpha(t)$ combine hadrons into families:
 $s_h=\alpha(m_h^2)$, where $s_h$ and $m_h$ are the spin and the mass
 of a hadron (hadronic resonance) with given quantum numbers (baryon
 number, isotopic spin, strangeness, etc.)~\cite{CF}. Moreover, at
 negative values of $t$, that is in the physical region of the
 $s$-channel, the very same function $\alpha(t)$ determines the
 scattering amplitude, according to Eq.~\eqref{Fst}.  It looks {\em as
 if}\/ high-energy scattering was due to $t$-channel exchange of a
 ``particle'' with spin $\alpha(t)$ that varies with momentum transfer
 $t$ --- the ``reggeon''.

 Thus, the high-energy behaviour of the scattering process $a+b\to
 c+d$ is linked with the spectrum of excitations (resonances) of
 low-energy scattering in the dual channel, $a+\bar{c}\to \bar{b}+d$.
 This intriguing relation triggered many new ideas (bootstrap, the
 concept of duality).
 Backed by the mysterious {\em linearity}\/ of Regge trajectories
 relating spins and squared masses of observed hadrons, the duality
 ideas, via the famous Veneziano amplitude, gave rise to the concept
 of hadronic strings and to development of string theories in general.

\subsection{Interacting Pomerons}

 A lot of theoretical effort was invested into understanding of the
 approximately constant behaviour of total cross sections at high
 energies.

 To construct a full theory that would include the Pomeron trajectory
 with the maximal ``intercept'' $\alpha_P(0)\!=\!1$ that respects the
 Froissart bound, and would be consistent with unitarity and
 analyticity proved to be very difficult.
 This is because multi-Pomeron exchanges become essential, which
 generate {\em branch points}\/ in the complex $\ell$-plane,
 which singularities were first discovered by Mandelstam in his
 seminal paper~\cite{Mcuts}.
%
%
 Moreover, the study of particle production processes with large
 rapidity gaps led 
 Gribov, 
 Pomeranchuk and
 Ter-Martirosyan to the concept of {\em interacting reggeons}.
 By the end of the 60-s V.~Gribov had proposed the general theory
 known as Gribov Reggeon Calculus. He formulated the rules for
 constructing the field theory of interacting Pomerons --- the
 Reggeon Field Theory (RFT) --- and developed the corresponding
 diagram technique.  Gribov RFT reduces the problem of high energy
 scattering to a non-relativistic QFT
 of interacting particles in 2+1 dimensions.

\subsection{Gribov Partons and Feynman Partons}

 One of Gribov's most important contributions to high energy hadron
 physics was the understanding of the space-time evolution of high
 energy hadron-hadron and lepton-hadron processes, in particular the
 nature of the reggeon exchange from the $s$-channel point of view.
 In 1973, in his lecture at the LNPI Winter School~\cite{lecture},
 Gribov outlined the general phenomena and typical features that were
 characteristic for high energy processes in any QFT.
%

 To understand the structure of hard (deep inelastic) photon--hadron
 interactions Feynman suggested the idea of partons --- point-like
 constituents of hadronic matter.  Feynman defined partons in the
 infinite momentum frame to suppress vacuum fluctuations whose
 presence would have undermined the notion of the parton wave function
 of a hadron~\cite{Feynm}.

 The power of Gribov's approach lied in applying the universal picture
 of fluctuating hadrons to both {\em soft}\/ and {\em hard}\/
 interactions.
 Gribov's partons are constituents of hadron matter, components of
 long-living fluctuations of the hadron projectile, which are
 responsible for soft hadron-hadron interactions: total cross
 sections, diffraction, multiparticle production, etc.

 Gribov's earlier work {\em Interaction of $\gamma$-quanta and
 electrons with nuclei at high energies}~\cite{gammanucl} had been a
 precursor to the famous Feynman paper.  Gribov described the photon
 interaction in the rest frame of the target nucleus.  An incident
 real photon (or a virtual photon in the deep inelastic scattering
 case) fluctuates into hadrons before the target, at the longitudinal
 distance $L$ increasing with energy.\footnote{For the $e^-p$ deep
 inelastic scattering B.L.~Ioffe has shown that the assumption of
 Bjorken scaling implies $L\sim 1/2xm_N$, with $x$ the Bjorken
 variable and $m_N$ the nucleon mass~\cite{Ioffe}.}
  Therefore, at sufficiently large energy, when the fluctuation
  distance exceeds the size of the target, the photon no longer
  behaves as a point-like weekly interacting particle. Its interaction
  resembles that of a hadron and becomes ``black'', corresponding to
  complete absorption on a large nucleus.
%

 Being formally equivalent to Feynman's treatment, Gribov's approach
 is better suited for the analysis of deep inelastic phenomena at very
 small Bjorken $x$, where the interaction becomes actually strong, and
 the perturbative QCD treatment is bound to fail.
 Gribov diffusion in the impact parameter space giving rise to energy
 increase of the interaction radius and to the reggeon exchange
 amplitude, coexisting fluctuations as a source of branch cuts,
 duality between hadrons and partons, a common basis for hard and soft
 elastic, diffractive and inelastic process --- these are some of the
 key features of high energy phenomena in QFTs,
 which are still too hard a nut for QCD to crack.

\subsection{Gribov Reggeon Field Theory}

The two best known applications of the Gribov RFT are
\begin{itemize}
\item
 general quantitative relation between the shadowing phenomenon in
 hadron-hadron scattering, the cross section of diffractive processes
 and inelastic multi-particle production, known as
 ``Abramovsky-Gri\-bov-Kan\-chelli cutting rules'' (AGK)~\cite{AGK},
 and
\item 
 the essential revision by Gribov of the Glauber theory of nuclear
 shadowing in hadron-nucleus interactions~\cite{GGlaub}.
\end{itemize}
 In 1968 V.N.\ Gribov and A.A.~Migdal demonstrated that the scaling
 behaviour of the Green functions emerged in 
QFT
 in the strong coupling regime~\cite{GM}. As we have mentioned in the
 Introduction, their technique helped to build the quantitative theory
 of second order phase transitions and to analyse critical indices
 characterising the long-range fluctuations near the critical point.

 The problem of high energy behaviour of soft interactions remained
 unsolved, although some viable options were suggested. In particular,
 in {\em Properties of Pomeranchuk Poles, diffraction scattering and
 asymptotic equality of total cross sections}~\cite{equal} Gribov has
 shown that a possible consistent solution of the RFT in the so-called
 {\em weak coupling}\/ regime calls for the formal asymptotic equality
 of {\em all}\/ total cross sections of strongly interacting
 particles.

 Gribov's last work in this subject was devoted to the intermediate
 energy range and dealt with interacting hadron
 fluctuations~\cite{heavyP}.

 The study of the {\em strong coupling}\/ regime of interacting
 reggeons (pioneered by A.B.\ Kai\-da\-lov and K.A.~Ter-Martirosyan)
 led to the introduction of the {\em bare}\/ Pomeron with
 $\alpha_P(0)\!>\!1$. The RFT based on $t$-channel unitarity should
 enforce the $s$-channel unitarity as well. The combination of
 increasing interaction radius and the amplitudes in the impact
 parameter space which did not fall as $1/\ln s$ (as in the
 one-Pomeron picture) led to logarithmically increasing asymptotic
 cross sections, resembling the Froissart regime (and respecting the
 Froissart bound Eq.~\eqref{stot}).
 The popularity of the notion of the bare Pomeron with
 $\alpha_P(0)\!>\!1$ is based on experiment (increasing 
 $\sigma_{\mbox{\scriptsize tot}}$).
 Psychologically, it is also supported by the perturbative QCD finding
 that the (small) scattering cross section of two
 small-transverse-size objects should increase with energy in a
 power-like fashion in the restricted energy range --- the famous
 ``hard BFKL Pomeron''~\cite{BFKL}.

\subsection{Reggeization and 
Pomeron Singularity in Gauge Theories}

 In the mid-sixties Gribov initiated the study of double logarithmic
 asymptotics of various processes in QED, making use of the powerful
 technique he had developed for the analysis of the asymptotic
 behaviour of Feynman diagrams~\cite{Gribov_lectures}.

 In particular, in 1975 
 Gribov,
 Lipatov and 
 Frolov studied the high energy behaviour of QED processes from the
 point of view of ``Regge theory''. High energy scattering amplitudes
 with exchange of an {\em electron}\/ in the $t$-channel acquire, in
 higher orders in QED coupling, a characteristic behaviour $A\propto
 s^{j(t)}$ with $j(m_e^2)=1/2$. This means that electron becomes a
 part of the Regge trajectory: {\em reggeizes}.\footnote{And so do
 quarks and gluons in QCD;
Fadin, 
Frankfurt, 
Lipatov,
Sherman (1976).}  
 For the {\em vacuum channel}, however, they
 found~\cite{qedjplane} that the rightmost singularity in the complex
 $j$-plane is {\em not a moving pole}\/ (as it is for electron) but,
 instead, {\em a fixed branch point}\/ singularity positioned {\em to
 the right}\/ from unity, $j(0)=1+c\alpha^2>1$.
 This was a precursor of a similar result found later by Fadin,
 Lipatov and Kuraev~\cite{BFKL} in non-Abelian theories, and QCD in
 particular.  The problem of apparent anti-Froissart behaviour of the
 perturbative ``hard Pomeron'' in QCD still awaits resolution.

 With the advent of QCD as a microscopic theory of hadrons and their
 interactions, the focus of theoretical studies has temporarily
 shifted away from Gribov--Regge problematics to ``hard''
 small-distance phenomena.

\section{ 
NON-ABELIAN GAUGE THEORIES}

 V.N.~Gribov became interested in non-Abelian field theories in
 1976. His very first study, as a QCD apprentice, produced amazing
 results.
 In February 1977 in the proceedings of the 12$^{\mbox{\scriptsize th}}$ Leningrad Winter
 School he published two lectures which were to change forever the
 non-Abelian landscape~\cite{BH1,BH2}.

\subsection{Anatomy of Asymptotic Freedom}

 In the first lecture {\em Instability of non-Abelian gauge fields and
 impossibility of the choice of the Coulomb gauge}~\cite{BH1} Gribov
 gave an elegant physically transparent explanation of 
 the {\em asymptotic freedom}\/ introduced by D.\ Gross \& F.\ Wilczek
 and H.D.\ Politzer in 1973~\cite{AsFr}.

 The {\em anti-screening}\/ phenomenon had been first observed 
 back in 1969~\cite{Julik} for the non-Abelian $SU(2)$ theory in the
 ghost-free Coulomb gauge within the Hamiltonian approach.  In the
 Hamiltonian language, there are (or rather seem to be) $N^2\!-\!1$
 massless ${\bf B}$ quanta (transverse gluons, $\bperp$) and, as in
 QED, an additional Coulomb field (${\bf 0}$). It is important to
 stress that the latter is not a physical quantum degree of freedom
 but
a means
 for describing classical
 instantaneous interaction between colour charges. Unlike QED, the
 non-Abelian Coulomb field has a colour charge of its own. Therefore,
 traversing the space between two external charge, it may virtually
 decay into two transverse fields,
\bminiG{eq:qcddec}
   {\bf 0} \>\to\> \bperp + \bperp \>\to\> {\bf 0}\,, 
\end{eqnarray}
or into a $q\bar{q}$ pair 
\begin{eqnarray}
   {\bf 0} \>\to\> {\bf q} + \bar{\bf q} \>\to\> {\bf 0}\,, 
\emini
 in the same manner as the QED Coulomb field fluctuates in the vacuum
 into an $e^+e^-$ pair.  Both these effects lead to {\em screening}\/
 of the colour charge of the external sources, in a perfect accord
 with one's physical intuition.

 This ({\em anti-asymptotically-free}\/) behaviour of the running
 coupling was first found by Landau, Abrikosov and Khalatnikov for
 QED~\cite{LAK}. It turned out to be common for all then-known
 renormalizable field theories: with scalar ($\lambda\phi^4$), Yukawa,
 four-fermion interactions~\cite{Pom} as well as for pedagogically
 valuable Lee model~\cite{KP}.

 This observation had dramatic consequences: the physical interaction
 (renormalized coupling) was predicted to {\em vanish}\/ in the limit
 of a point-like bare interaction, $\Lambda_{\mbox{\scriptsize
 UV}}\to\infty$.  In the late 1950s the problem was known as ``Moscow
 Zero''.
 The depth of the crisis can be measured by the Dyson
 prophecy~\cite{Dyson} that the correct ``meson'' theory -- the theory
 of strong interactions -- ``{\em will not be found in the next
 hundred years}\/'' and by the Landau conclusion~\cite{Landaulast}
 that ``{\em the Hamiltonian method for strong interactions is dead
 and must be buried, although of course with deserved honour.}''


 Universality of the screening phenomenon was readily understood as a
 consequence of {\em unitarity}\/ in the cross-channel.
%
%
%
 Indeed, the discontinuity of the loop diagram at $t>0$, 
$$ 
\begin{picture}(0,0)%
\includegraphics{discon.pstex}%
\end{picture}%
\setlength{\unitlength}{3947sp}%
\begingroup\makeatletter\ifx\SetFigFont\undefined%
\gdef\SetFigFont#1#2#3#4#5{%
  \reset@font\fontsize{#1}{#2pt}%
  \fontfamily{#3}\fontseries{#4}\fontshape{#5}%
  \selectfont}%
\fi\endgroup%
\begin{picture}(4072,762)(136,-151)
\end{picture}
 \qquad \propto \quad \sigma\left(\>\> \begin{picture}(0,0)%
\includegraphics{sigma.pstex}%
\end{picture}%
\setlength{\unitlength}{3947sp}%
\begingroup\makeatletter\ifx\SetFigFont\undefined%
\gdef\SetFigFont#1#2#3#4#5{%
  \reset@font\fontsize{#1}{#2pt}%
  \fontfamily{#3}\fontseries{#4}\fontshape{#5}%
  \selectfont}%
\fi\endgroup%
\begin{picture}(642,644)(294,-75)
\end{picture}
 \right),
$$ 
 describes, according to Cutkosky rules, production cross section of
 two physical ({\em sic}\/!)  particles in the $t$-channel. Positivity
 of the imaginary part at $t>0$ then dictates the {\em screening}\/
 sign of the virtual loop in the $s$-channel ($s>0$, $t<0$).

 Thus, the fact that there are ``physical'' fields in the intermediate
 state in Eq.~\eqref{eq:qcddec} --- quark-antiquark or two transverse
 gluons --- fixes the sign of the virtual correction to correspond to
 {\em screening}, via the unitarity relation in the cross-channel.
 Technically speaking, these virtual decay processes contribute to the
 QCD $\beta$-function as
\bminiG{beta}
\label{eq:betaphys}
\left\{ \frac{d\,\alpha_s^{-1}(R)}{d\ln R}\right\}^{\mbox{\scriptsize phys}}  
\>\propto\>\> \frac13\, N \,+\,  \frac{2}{3}\,n_f\,,
\end{eqnarray}
 that is, make the effective coupling {\em decrease}\/ at large
 distances $R$ between the external charges, as in QED.

 Where then the {\em anti-screening}\/ comes from?  It originates from
 another, specifically non-Abelian, effect namely, interaction of the
 Coulomb field with the field of ``zero-fluctuations'' of transverse
 gluons in the vacuum,
\[
\sum_n  \Big[ {\bf 0}\>+ \bperp \>\to\> {\bf 0}\Big]^n \quad =  \qquad
\begin{picture}(0,0)%
\includegraphics{coulomb.pstex}%
\end{picture}%
\setlength{\unitlength}{3947sp}%
\begingroup\makeatletter\ifx\SetFigFont\undefined%
\gdef\SetFigFont#1#2#3#4#5{%
  \reset@font\fontsize{#1}{#2pt}%
  \fontfamily{#3}\fontseries{#4}\fontshape{#5}%
  \selectfont}%
\fi\endgroup%
\begin{picture}(3037,674)(879,-113)
\end{picture}
 + \ldots .  
\]
 In a course of such multiple rescattering, the Coulomb ``quantum''
 preserves its identity as an {\em instantaneous}\/ interaction
 mediator, and therefore is not affected by the unitarity constraints.
 For $n=1$ the contribution vanishes upon averaging. Statistical
 average over the transverse vacuum fields in the second order of
 perturbation theory ($n=2$) results in an additional contribution to
 the Coulomb interaction energy which, translated into the running
 coupling language, gives
\begin{eqnarray}
\label{eq:betastat}
\left\{ \frac{d\,\alpha_s^{-1}(R)}{d\ln R}\right\}^{\mbox{\scriptsize stat}}  
\>\propto\> -4\, N\,. 
\emini
 According to Gribov, an anti-intuitive minus sign in
 Eq.~\eqref{eq:betastat} has its own simple explanation. It is of the same
 origin as the minus sign in the shift of the energy of the ground
 state of a quantum-mechanical system under the second order in
 perturbation:
\[
 \delta E \equiv E-E_0 
 = \sum_n \frac{\abs{\lrang{0|\delta V|n}}^2}{E_0-E_n} \> < \> 0\,.
\] 
 The r\^ole of perturbation $\delta V$ is played 
 by the vacuum field of transverse gluons.

 Taken together, the two contributions Eq.~\eqref{beta} combine into the
 standard QCD $\beta$-function:
\[
   \frac{d}{d\ln Q^2} \frac{\alpha_s(Q^2)}{4\pi} =
   -b_0\left(\frac{\alpha_s(Q^2)}{4\pi}\right)^2 + \cO{\alpha_s^3},
\qquad b_0 = \frac{11N_c}{3} - \frac{2N_f}{3}.
\]

\subsection{Infrared Instability}

 The three-dimensional transversality condition
\begin{equation}
\label{eq:Btr}
 \left(
\bnabla
\cdot {\bf A}\right) 
  \equiv   \frac{\partial {A}_i}{\partial x_i } =
  0\,, \qquad i=1,2,3
\end{equation}
 is usually imposed on the field potential to describe massless vector
 particles (the Coulomb gauge). 
 Being antisymmetric, the field strength tensor $F^a_{\mu\nu}$ does
 not contain {\em time derivative}\/ of the zero-component of the
 potential ${A}^a_0$.  In the Hamiltonian language, this puts
 ${A}^a_0$ in a position of a {\em cyclic variable}\/ which does not
 constitute a physical degree of freedom.  It can be eliminated from
 field dynamics~\cite{BH1,ChristLee},
 contributing to the Hamiltonian that describes transverse gluons,
\bminiG{Hamil}
\label{Hglue} 
 H_{\mbox{\scriptsize glue}} = \oh \int d^3x\;\left( \J^{-\oh}\, {\bf
 E}_\perp^{a}\> {\cal J} \cdot {\bf E}_\perp^{a} \J^{-\oh} \>+\> {\bf
 B}_\perp^a \cdot {\bf B}_\perp^a \right),
\end{eqnarray} 
 an additional term responsible for 
 {\em Coulomb interaction}\/ between ``charges'',
\begin{eqnarray}
\label{HCoul} 
 H_{\mbox{\scriptsize Coul}} = \oh \int d^3x\, d^3y\;\J^{-\oh}\rho^a(x)\, \J\>
 K^{ab}(x,y;{\bf A}_\perp)\> \rho^b(y)\, \J^{-\oh} ,
\end{eqnarray}
 where $\rho^a$ is given by the sum of the colour charge density of
 external sources (e.g., static quarks) and that induced by the
 transverse gluons themselves,
\begin{eqnarray}
\label{Hdensity} 
 \rho
 \>=\> \rho_q
 \>+\> i g_s\, 
 [ {\bf  A}_\perp, {\bf E}_\perp ].
\emini
Introducing the covariant derivative ${\bf D}$,
\[
 {\bf D}\,[{\bf A}
 ]\, C \>=\> \bnabla\,C + ig_s\left[{\bf A}
 ,C\right],
\]
the Coulomb energy ``propagator'' $K$ in Eq.~\eqref{HCoul} reads
\begin{equation}
\label{eq:Kdef}
 K^{ab}(x,y;{\bf A}) 
\>=\> - \left[ {1\over {\bf D}[{\bf A}]\cdot  \bnabla }
      \> \bnabla^2 \> {1\over {\bf D}[{\bf A}]\cdot  \bnabla }
       \right]^{ab}_{xy}.
\end{equation}
 We see that the propagation of a Coulomb field $A_0
$ in the ``external'' transverse
 gluon field ${\bf A}
_\perp$
 is governed by the operator
 that resembles the propagator of the Faddeev--Popov ghost,
\begin{equation}
\label{eq:FPgh}
 ({\bf D}\cdot\bnabla )\, A_0 \>\equiv\> \bnabla^2 A_0 +
 ig_s\left[{\bf A}_{\perp}, \bnabla A_0
\right].
\end{equation}
 The factor $\J$ in Eq.~\eqref{Hamil} is the determinant of this
 operator,
\[
 \J=\J[{\bf A}] \>=\> -\det\left( {\bf D}[{\bf A}] \cdot \bnabla\right).
\]
 Taking the expectation value of $K(x,y;{\bf A})$ over the vacuum
 fields ${\bf A}_\perp$ produces an {\em instantaneous}\/ interaction term,
\begin{eqnarray}
         \left\langle A^a_0(x) A^b_0(y) \right\rangle &=&
  G(\bx-\by)\d^{ab} \d(x_0-y_0) \non \\
            &+& \mbox{non-instantaneous},
\non \\
      G(\bx-\by)\,\d^{ab} &=& -\left\langle \left[{1\over {\bf D}[{\bf
      A}_\perp]\cdot \bnabla} \>\bnabla^2\> {1\over {\bf D}[{\bf
      A}_\perp]\cdot \bnabla} \right]_{x,y}^{a,b}
\right\rangle.
\label{D}
\end{eqnarray}
 This expression describes an interaction generated by exchange of a
 Coulomb gluon {\em dressed}\/ by the fluctuations of transverse gluon
 fields in the vacuum.

 By setting ${\bf A}_\perp=0$ we would return to the Laplace operator
 $G=-1/\bnabla^2 \propto 1/{\bf k}^2$ that corresponds, in the
 coordinate space, to the canonical Coulomb potential
 $1/\abs{\bx-\by}$.
 For small vacuum fields, $g_s{A}_\perp/\nabla\ll 1$, expanding
 perturbatively the Coulomb propagator $G$ in Eq.~\eqref{D} to the
 second order in $g_s$ produces the one-loop anti-screening effect as
 stated in Eq.~\eqref{eq:betastat}.

 If we take, however, gluon fields in the QCD vacuum as large as
\[
  g_s{ A}_\perp/\nabla \sim g_s\,{ A}_\perp\cdot L \sim 1
\] 
 (with $L$ a spatial extent of the field), the perturbative expansion
 in $g_s{\bf A}_\perp$ of the denominators in Eq.~\eqref{D} is no
 longer justified. Under such circumstances a qualitatively new
 phenomenon takes place namely, the Coulomb (ghost) propagator may
 become singular:
\begin{equation}
\label{eq:zero}
\left( {\bf D}[{\bf A}_\perp]\cdot \bnabla\right)\, C_0 \>=\> \bnabla^2 C_0 +
      ig_s\left[\,{\bf A}_{\perp},\bnabla C_0 \,\right] \>=\>0\,.
\end{equation}
 Appearance of a ``zero-mode'' solution $C_0$ in the external field is
 a sign of an infrared instability of the theory.

 An illuminating way to see the instability of the perturbative
 vacuum, and to re-derive the running of the coupling in QCD, is to
 look at the quantum correction to the vacuum energy $V_0(H)=\oh H^2$
 in a constant chromo-magnetic field $H$.

 The one-loop corrected energy density reads~\cite{MatSav}
\beq
 \Re V(H) = \frac{1}{2} H^2 + (g_sH)^2 \frac{b}{32 \pi^2} \left( \ln
 \frac{g_s H}{\mu^2} - \frac{1}{2} \right) \>\simeq\>
 \frac{g_s^2(\mu^2)}{g_s^2(H)}\cdot V_0(H)
, \label{MS}
\eeq
 where $\mu$ is the renormalization scale of the bare coupling,
 $g_s=g_s(\mu^2)$, and $b = 11 N_c / 3$ (gluodynamics).  
 For relatively small fields,
\[
  H \> < \> H_0 \simeq \frac{\mu^2}{g_s(\mu^2)} \exp \left(- \frac{16
  \pi^2}{b g^2(\mu^2)}\right),
\]
 the potential Eq.~\eqref{MS} formally turns {\em negative}\/ and
 develops a minimum --- ``true vacuum'' (?) --- corresponding to
 non-zero expectation value of the chromo-magnetic field.  It
 was soon realized, however, that this new ``vacuum'' is
 unstable~\cite{NO78}.

 In pure gluodynamics the vacuum correction can be computed by summing
 over the Landau levels of gluons moving in the external field:
\beq
\delta V(H) = \frac{g_s H}{4 \pi^2} \ \int d p_z \sum_{n=0}^{\infty}
 \sum_{s_z=\pm 1} \sqrt{ 2g_s H (n + \oh - s_z) + p_z^2}.
 \label{Landaulev}
\eeq
 The real part of (\ref{Landaulev}) yields (\ref{MS}).  Thus, the
 asymptotic freedom in this approach can be seen as arising from the
 paramagnetic response of the vacuum. 

 At the same time, we readily see that one specific state in
 Eq.~\eqref{Landaulev} namely, $n=0$ with gluon spin parallel to the
 direction of the field, $s_z = 1$, gives an {\em imaginary}\/
 contribution:
\beq
  \Im \delta V(H) = \frac{g_sH}{4 \pi^2} \int_{-g_sH}^{g_sH} d p_z
  \Im  \sqrt{p_z^2 - g_sH - i0} = - \frac{g_s^2 H^2}{8 \pi} =
  -\frac{g_s^2}{4 \pi} \cdot V_0(H).
\eeq 
 Non-zero imaginary part of the effective potential signals that the
 vacuum is unstable (decays with time) and does not correspond to the
 true ground state of the theory. Note also that $n=0$ corresponds to
 the Landau level of the largest transverse radius, i.e.\ to the {\em
 infrared}\/ region, as had to be expected.

 Once again, we come to the conclusion that \naive\ perturbative
 treatment of non-Abelian gauge fields is flawed due to problems in
 the infrared.

\subsection{Gribov Copies}

 In \cite{BH1} Gribov localized the problem. He showed that the
 three-dimensional transversality condition Eq.~\eqref{eq:Btr}
 actually {\em does not}\/ solve the problem of gauge fixing. 

 Consider a group $\Omega$, and let $\omega(x)$ be a function with
 values in this group. The basic principle that defines the
 corresponding gauge theory is that the vector potentials $A_{\mu}(x)$
 and
\beq \label{nonl}
 A_{\mu}^{[\omega]}(x) = \omega(x) A_{\mu}(x) \omega^{-1}(x) +
 \partial_{\mu} \omega(x) \omega^{-1}(x)
\eeq
 describe physically identical fields. Therefore, to avoid multiple
 counting,
 one has to impose a gauge-fixing condition of the general form
\beq
 F(A^{[\omega]}; x) = 0. \label{gaugefix}
\eeq
 To do the job, 
 Eq.~\eqref{gaugefix} should have a unique solution for arbitrary
 $A$. This can be achieved for a variety of gauge fixings
 within perturbation theory, when the fields are {\em weak}. In
 general, however, solutions may appears copious since the gauge
 fixing condition Eq.~\eqref{gaugefix} constitutes a system of
 nonlinear equations for the function $\omega(x)$.

 Indeed, Gribov found that due to essential non-linearity of the gauge
 transformation Eq.~\eqref{nonl}, a ``transverse'' field potential
 satisfying Eq.~\eqref{eq:Btr} may actually happen to be a {\em pure
 gauge field}\/ which should not be separately counted as an
 additional physical degree of freedom.
 He explicitly constructed such ``transverse gauge fields'' for the
 $SU(2)$ gauge group and showed that the {\em uncertainty}\/ in gauge
 fixing arises when the effective magnitude of the field becomes
 large,
\[
     A_\perp \cdot L \>\sim \frac{1}{g_s},
\]
 or, in other words, when the effective interaction strength (QCD
 coupling) becomes of the order of unity, that is, in the
 non-perturbative region.
 More precisely, he found that it happens exactly when the
 Faddeev--Popov operator acquires a zero mode solution
 Eq.~\eqref{eq:zero} that is, as we discussed above, in the infrared
 region where the vacuum enhancement of the dressed Coulomb gluon
 propagator (\ref{D}) becomes catastrophically large.

 Thus, the ``surface'' $({\bf D}[{\bf A}_\perp]\cdot\bnabla) C_0=0$ in
 the functional ${\bf A}_\perp$--space
 marks the border (the ``Gribov horizon'') beyond which the solution
 of the gauge-fixing equation Eq.~\eqref{eq:Btr} becomes copious.
 From this point of view, the fact that the Coulomb propagator
 develops singularity does not necessarily mean that the
 Faddeev--Popov ghost ``rises from the dead'' by pretending to
 propagate as a particle. It rather tells us that we have failed to
 formulate the quantum theory of interacting non-Abelian vector
 fields, to properly fix physical degrees of freedom.

 The existence of ``Gribov copies'' means that the standard
 Faddeev--Popov prescription for quantizing non-Abelian gauge theories
 is, strictly speaking, incomplete and has to be modified.


 A number of studies explored further the emergence of Gribov
 copies. It was found, both analytically \cite{Semenov} and
 numerically \cite{num} that Gribov copies can exist even {\em
 inside}\/ the Gribov horizon, and that a more narrow ``fundamental
 modular region'' had to be defined to avoid the
 problem. Nevertheless, recently it has been shown \cite{Zwan04} that
 the copies within the Gribov horizon 
 actually 
 do not contribute to any expectation values, and thus for all
 practical purposes the original Gribov's recipe of constraining
 Faddeev-Popov determinant to positive values is correct.
 A promising 
 attempt 
 to implement the Gribov fundamental domain in the 5-dimensional
 formulation of gluodynamics can be found in \cite{BZ}.

 \subsection{Coulomb Confinement}

 Gribov 
 himself 
 addressed the 
quest
 of possible modification of the QCD quantization procedure in the
 second lecture ``Quantization of non-Abelian gauge
 theories''~\cite{BH2}.
 The paper ~\cite{Quant} under the same title based on the two Winter
 School lectures is now a universally accepted (though disturbing)
 truth and during 25 odd years since its appearance in 1978 was being
 cited more than
 660 times, with increasing frequency.

 To properly formulate non-Abelian field dynamics, Gribov suggested to
 limit the integration over the fields in the functional integral to
 the so-called {\em fundamental domain}, where the Faddeev--Popov
 determinant is strictly positive (the region in the functional space
 of transverse fields ${\bf A}_\perp$ {\em before}\/ the first zero
 mode Eq.~\eqref{eq:zero} emerges).

 Gribov produced qualitative arguments in favour of the characteristic
 modification of the gluon propagator, due to the new restriction
 imposed on the functional integral. Effective suppression of large
 gluon field results, semi-quantitatively, in an infrared singular
 polarization operator $\Pi\propto k^{-2}$, 
\bminiG{Gprop}
\label{eq:Gprop1}
 D^{-1}(k) = k^2 +\Pi(k^2) \>\simeq\> k^2 + \frac{{\sigma^2}}{k^2}.
\end{eqnarray}
 The gluon Green function coincides with the perturbative one
 at large momenta (small distances), $D(k)\propto k^{-2}$ but {\em
 vanishes}\/ off at $k=0$, instead of having a pole corresponding to
 massless gluons:
\begin{eqnarray}
\label{eq:Gprop2}
  D(k) \propto \frac{k^2}{k^4 + {\sigma^2}} . 
\emini
 The new non-perturbative parameter $\sigma^2$ in Eq.~\eqref{Gprop}
 has dimension (and the meaning) of the familiar vacuum gluon
 condensate, $\sigma^2\sim \lrang{\alpha_s (F^a_{\mu\nu})^2}$, that
 emerged in the context of QCD sum rules \cite{SVZ}. 

 Literally speaking, the ansatz Eq.~\eqref{eq:Gprop2} cannot be
 correct since such a Green function would violate
 causality.\footnote{It is unfortunate, therefore, that the form
 Eq.~\eqref{eq:Gprop2} which Gribov suggested and discussed for
 illustrative purposes only, is often referred to in the literature as
 ``the Gribov propagator''.}
 In reality, the gluon (as well as the quark) propagator should have a
 more sophisticated analytic structure with singularities on
 unphysical sheets, which would correspond, in the standard
 field-theoretical language, to {\em unstable}\/ particles.

 At the same time, the modification of the Coulomb (ghost) propagator
 due to restricting the functional integral to the fundamental domain
 \cite{BH2} resulted in the singular small momentum behaviour
\[
    G(k) \propto \frac{1}{N_c\,g_s^2}\cdot \frac{\sigma}{{\bf k}^4},
\]
 corresponding to a linear increase of the interaction energy 
 at large distances $R=\abs{\bx-\by}$ between colour charges,
 $V(R)\propto\sigma R$.


 The idea of confinement emerging from dressed Coulomb exchange was
 further explored by Zwanziger \cite{Zwan98} who has recently shown
 {\cite{Zwan03} that for the static interaction potential $V(R)$
 the following inequality holds:
\beq\label{Zwineq}
    V(R) \leq V_{\mbox{\scriptsize Coul}}(R).
\eeq
 This means that if confinement exists in pure gluodynamics, it should
 arise already at the dressed one-gluon exchange level --- ``{\em no
 confinement without Coulomb confinement}\/'' \cite{Zwan03}.  
%
 The inequality Eq.~\eqref{Zwineq} appears inevitable: an inclusion of
 ``quantum'' (gluons, quarks) degrees of freedom can only {\em
 screen}, that is, suppress, the classical (Coulomb) interaction, as
 we have discussed in the beginning of this Section.
 
 Recent lattice studies of the correlator of timelike link variables
 in Coulomb gauge \cite{lattice} show that the Coulomb interaction
 energy of static quarks indeed grows linearly
 at large quark separations.  
 This simulation was designed to measure the {\em pure}\/ Coulomb
 energy, without admixture of additional ``constituent'' gluons,
\beq\label{pure_string}
 \Psi \>=\> \bar{q}^{a}({\bf x})\ q^{a} ({\bf y}) \Psi_0.
\eeq 
 The corresponding string tension $\sigma$ turned out to be several
 times greater than the generally accepted value that originated from
 lattice measurements of the {\em full}\/ static interaction energy.
 
 This fact is remarkable. It shows that the Coulomb exchange excites
 transverse vacuum gluons inclusion of which results in the {\em
 energy gain}. These gluons ``shake'' the initial string
 Eq.~\eqref{pure_string} and {\em soften}\/ it.

 In this picture the typical transverse size of the string grows {\em
 logarithmically}\/ with quark separation (the so-called
 ``roughening''). 
 It is worthwhile to notice that this phenomenon is similar to the
 increase of the size of a hadron at high energies as described by
 ``Gribov diffusion'', multiplication of the number of gluons being
 the physical reason for both.

 \subsection{Perturbative (Gluon) Confinement}

 Motivated by Gribov ideas, J.~Greensite and C.~Thorn recently
 suggested and developed an interesting model that aims at reproducing
 the main features of the Coulomb confinement essentially {\em
 perturbatively}~\cite{GreenThorn}.


 They proposed the model in which gluons are arranged in {\em
 chains}\/ that stretch between external charges.  The Fock state wave
 function of such a chain
 looks like
\beq\label{chain_string}
 \Psi_{chain}[A] = \bar{q}^{a_1}({\bf x})\ A^{a_1 a_2}({\bf x_1})\
 A^{a_2 a_3}({\bf x_2})\> \ldots\> A^{a_N a_{N+1}}({\bf x_N}) \ q^{a_{N+1}}
 ({\bf y})\, \Psi_0[A].
\eeq 
 The spatial distribution of the gluons is then determined by
 minimizing the total energy. 

 Qualitative explanation of the Coulomb confinement in terms of ``the
 gluon chain model'' of Greensite and Thorn runs as follows.
 In 't Hooft's large--$N_c$ limit ($N_c \to \infty$, $g_s^2 N_c$
 fixed), the dominant diagrams are the planar ones in which each of
 the gluons interacts only with its nearest neighbors. (Note that the
 nearest-neighbor interaction can be purely Coulombic, i.e.\
 perturbative.)  
 This state can be approximately represented (modulo $1/N_c^2$
 corrections) as a superposition of $(N+1)$ fundamental colour
 dipoles. As the distance between the quarks increases, so does the
 number of gluons (dipoles) $N$ between them, so that the number of
 gluons per unit length
%
 turns out
 to stay finite, $N/R = 1/\ell=$const, as 
 Monte Carlo 
 simulations showed. If $E_{\mbox{\scriptsize gluon}}$ is the kinetic
 plus nearest-neighbor interaction energy of each gluon, then the
 total energy of the system is
\beq
 E(R) = N E_{\mbox{\scriptsize gluon}} = \frac{E_{\mbox{\scriptsize
 gluon}}}{\ell}\ R \equiv \sigma R,
\eeq
 where the the r\^ole of the string tension is played by gluon energy
 per unit length, $\sigma = E_{\mbox{\scriptsize gluon}}/\ell$.

 It has been argued \cite{GreenThorn} that this model of confinement
 explains the Casimir scaling of the string tension observed on the
 lattice (the proportionality of the confining force between static
 sources in representation $r$ of the gauge group to the quadratic
 Casimir operator $C_r$ of the representation) and yields the expected
 ``center dependence'' (the dependence of the string tension on the
 transformation properties of the center subgroup of the gauge group
 $\Omega$ --- ``$N$-ality'').

 Thus the linear growth of the interaction potential with distance is
 the consequence of the linear increase in the number of gluons
 excited by the static quark--antiquark interaction~\cite{GreenThorn}.
 It is important to stress that $N$ linearly increasing with distance
 in the {\em static}\/ problem, translates into the uniform
 rapidity distribution of excited gluons in the {\em relativistic}\/
 situation when external sources 
 fly
 away 
 with the speed of light as, for example, in electron-positron
 annihilation $e^+e^-\to q\bar{q}\to\mbox{hadrons}$.
 
 Such a picture therefore directly relates the notion of a nearly
 constant confining inter-quark force that is used to describe mass
 spectra of heavy quark bound states~\cite{linear}, with well
 developed ``string'' phenomenology of multiparticle production in
 high energy interactions~\cite{Lund_string}. The latter successfully
 describes the gross features of both soft and hard phenomena. On the
 {\em soft}\/ side we have ``multiperipheral'' hadroproduction in
 hadron--hadron reactions (Gribov plateau = Pomeron); on the {\em
 hard}\/ side --- the famous Feynman plateau, $xD(x)\simeq$const.,
 characteristic for parton distributions that explain spacelike deep
 inelastic scattering (and particle content of timelike jets) at small
 Bjorken (Feynman) $x$.

 This important link is not a monopoly of the gluon chain
 model. Indeed, we
 seem to be
 retelling the good old Kogut--Susskind confinement story: gluon field
 lines forming a string, digging up quark--antiquark pairs from the
 vacuum, a'la Schwinger tunneling mechanism
 in a constant field, assembling colourless hadrons in the final
 state. There is a crucial difference, however.
 We were used to look at
 the Kogut--Susskind scenario
 as calling for essentially non-perturbative dynamics: strong fields,
 ``superconductivity'' vortex picture to explain one-dimensional
 nature of the ``string'', etc. Now we are lead to think that the same
 physical picture can be achieved by purely {\em perturbative means},
 that is, by employing the language of the fundamental degrees of
 freedom of the theory --- gluons --- which gluons interact in a
 Coulomb-like (again, perturbative) fashion.

 't Hooft \cite{Hooft02} has advanced the Greensite--Thorn arguments,
 and pointed out that the ``gluon chains'' cannot provide a complete
 set of states 
 to confine the sources.
%
%
 The problem becomes apparent if we turn to the real world with light
 quarks present. Indeed, by pulling apart two heavy quarks we expect
 to find ourselves holding two colourless $D$-mesons, in the end of the
 day. Obviously, this cannot be achieved without including light
 vacuum quarks in the game. (Formally speaking, the adjoint
 representation gluons cannot fully blanch colour charges in the
 fundamental representation --- the external quarks.)
 Nevertheless, since gluon chain states appear to be energetically
 profitable, as recent lattice results have
 indicated~\cite{GreenThorn}, this model can serve as a good starting
 point, as a ``first approximation'' to the problem of quantitatively
 approaching the physics of hadrons.
 In particular, it may help to understand puzzling {\em softness}\/ of
 the transformation of partons into hadrons. These phenomena that one
 observed studying energy and angular distributions of soft particles
 produced in hard interactions known as ``local parton--hadron
 duality''~\cite{LPHD}, for recent reviews see \cite{KO,pylos}.

 't Hooft's (unpublished) remark \cite{Hooft02} bears an elegant
 strikingly simple title {\em Perturbative Confinement}.
 Looking back into historical perspective, we find this rather ironic,
 since this very endeavour --- perturbative approach to the
 confinement problem --- motivated the NPB referee to initially
 reject~\footnote{The second reason being, that ``the confinement
 problem had been already solved and isn't worth talking about.''}
 the pioneering Gribov paper \cite{Quant}.
 
 Instability of the perturbative vacuum tells us that in order to have
 a chance to approach the true {\em groung state}\/ of the theory by
 adiabatically switching on the interaction, one has to start from an
 {\em excited}\/ state, in terms of non-interacting perturbative
 vacuum. Implementation of this idea is being developed by P.~Hoyer
 and S.~Peigne, who are trying to model non-trivial structure of the
 vacuum by explicitly adding condensate terms to perturbative quark
 and gluon propagators~\cite{Hoyer}.


%

\subsection{OPE and Linear Potential}

 Another way of approaching the problem of confinement is to start at
 short distances, use the operator product expansion (OPE) and to
 study the structure of power corrections. For the case of potential
 acting between massive quarks inside a colourless bound state, the
 expansion takes the form \cite{hq}, see \cite{Zakharov03} for a
 recent survey,
\bea
  V(R) &\stackrel{R \to 0}{\simeq} & - 
%
 \frac{C_F\alpha_s(R)}{R} \left(1 + \sum_n a_n \alpha_s^n
 + c_3 \frac{\lrang{\alpha_s (F^a_{\mu\nu})^2}}{\alpha_s^2}\cdot R^4 \right)\!\!,
\label{instpot}
\eea
 where the second term in the brackets describes the higher order
 perturbative corrections, and the third one is the leading power
 correction. 
 The latter is motivated by the OPE that predicts the leading
 non-perturbative power correction to the vacuum energy to scale in
 accordance with the dimension of the vacuum expectation value of the
 first gauge invariant vacuum operator, $[\lrang{\alpha_s
 (F^a_{\mu\nu})^2}]=[R^{-4}]$.
 Within perturbation theory, its origin
 can be traced to ``infrared renormalons'', for a review see~\cite{renormalons}.



 What is most remarkable about the expression Eq.~\eqref{instpot} in
 the context of our discussion is that it does not contain a term
 linear in $R$ and thus contradicts the expected form
\beq
 V(R) \simeq - \frac{c}{R} + \sigma R,
\eeq
 borne out both by phenomenology of heavy quarkonia and by the lattice
 studies, as we have discussed above. 
 We therefore come to the troubling conclusion that the OPE is
 inconsistent with the linear confining potential.

 Yet another unexpected blow to the OPE ideology came from the direct
 state-of-the-art Monte Carlo simulation of the vacuum condensate in
 lattice gauge theory~\cite{1Q2} which showed that the leading
 non-perturbative correction to the Wilson loop plaquette expectation
 value scales with lattice spacing as $a^2$, instead of the
 OPE-blessed $a^{4}$.

 Various options were suggested to avoid this contradiction:
 (see~\cite{Zakharov03} and references therein).
 One can invoke new, genuinely non-perturbative, degrees of freedom as
 means for constructing missing dimension two vacuum condensates. The
 vocabulary of such approaches includes the notions of Dirac strings,
 (clustering) Monopoles, (percolating) Vortices and
 alike~\cite{Zakharov_Pom}.

 Alternatively, one may look at the QCD coupling itself; in other
 words, to search for $1/Q^2$ corrections emerging from the dressed
 gluon propagator.

\subsection{Infrared-Finite QCD Coupling}

 It is interesting that such terms naturally appear when one tries to
 implement the idea of {\em analyticity}\/ in the $Q^2$ variable by
 assuming the existence of spectral representation for the running
 coupling \cite{spectral,DMW}.
 For example, a $1/Q^2$-suppressed correction emerges as a result of 
 simple removal of the Landau pole from the one-loop
 coupling~\cite{Shirkov}:
\beq
 {\alpha_s(Q^2)} \>\Longrightarrow\> {\alpha_s}(Q^2) =
 \frac{4\pi}{b_0} \left(\frac{1}{\ln(Q^2/\Lambda_{QCD}^2)} +
 \frac{\Lambda_{QCD}^2} {\Lambda_{QCD}^2 - Q^2} \right). \label{anal}
\eeq
 The ansatz Eq.~\eqref{anal} may turn out to be too
 simplistic.\footnote{The sign of the $1/Q^2$ term it produces does
 not match that of the lattice result~\cite{Zakharov03}.}
 Nevertheless, the idea of enforcing analyticity (causality) on
 $\alpha_s
 $ 
 turned out to be very efficient, see in particular~\cite{gesh}. The
 corresponding technology that improves perturbative expansions is
 known under the name of ``Analytic Perturbation Theory''~\cite{APT}.
 Impressive recent results of refurbishing of perturbative series can
 be found in~\cite{CIPT}.

 A QCD couling that stays {\em finite}\/ all over the complex
 $Q^2$-plane (as an analytic $\alpha_s(Q^2)$ obviously does) allows
 one to peek into the strong interaction domain.
 One can aim at quantifying genuine confinement effects in QCD
 observables by linking power-suppressed non-perturbative
 contributions with momentum integrals (moments) of the coupling in
 the infrared region~\cite{DMW}.
 
 Markedly, the average value of the coupling that emerges from the
 study of the leading $1/Q$ power corrections to various jet shape
 observables in $e^+e^-$ annihilation and DIS turns out to be
\beq\label{a0}
 a_0\equiv \lrang{{\alpha_s(Q^2)}} = \frac{1}{2\GeV} \int_0^{2\GeV}
 dQ\, {\alpha_s(Q^2)}
 \>=\> 0.47\pm 0.07\,.
\eeq
 (For a recent review see \cite{pylos} and references therein.) 
 Let us mention, in passing, that this number turns out to be
 suspiciously close to the corresponding integral over
 Eq.~\eqref{anal}.  Even more interesting a ``coincidence'' is that
 the ``measured'' value Eq.~\eqref{a0} is comfortably above the
 so-called critical value of the strong coupling which is necessary,
 as we shall discuss in the next Section, to trigger the light quark
 confinement, according to Gribov, see Eq.~\eqref{eq:acrQCD} below.
 
 The quest for defining $\alpha_s$ at large distances has a long and
 turbulent history, for reviews see
 \cite{MatSt,incl,Brodsky}.
 On one hand, it goes without saying that the ``Landau pole'' is an
 artifact. On the other hand, in QCD we don't have means for defining
 the true ``physical coupling'' unambiguously, in contrast with QED
 where $\alpha(0)$ is directly accessible via small angle scattering,
 Josephson effect and alike.
 On the phenomenological side, Mattingly and Stevenson~\cite{MatSt}
 have assembled an impressive list of practical applications which
 consistently pointed at $\alpha_s/\pi = 0.2 - 0.3$ as a reasonable
 magnitude of the ``long-range'' QCD interaction strength.  The
 applications they considered ranged from rather \naive\ estimates of
 hadron-hadron cross sections and form factors to the well elaborated
 Godfrey--Isgur relativized QCD quarkonium model that described quite
 successfully particle spectroscopy from pions all the way up to the
 $\Upsilon$ family.  
 On the theory side, this quest ascends to the notion of ``effective
 charges'' introduced by G.~Grunberg back in 1984~\cite{effective}.
 In the context of the hunt for confinement effects, the concept of an
 infrared-finite coupling was suggested in \cite{hqalpha} for the
 purpose of quantifying non-perturbative $\Lambda_{QCD}/M_Q$ terms in
 fragmentation functions of heavy quarks. Perturbative {\em
 ideology}\/ behind this concept was laid down in~\cite{ideology}.

 The model coupling Eq.~\eqref{anal} {\em freezes}\/ at a constant
 value ($\alpha_s(0)=4\pi/b_0$). Though such a regime
 is often advocated in the literature,\footnote{for a recent emotional
 review see~\cite{Brodsky}} we'd rather it
 {\em vanished}\/ at the origin: any (even very weak) singularity at
 $k^2=0$ of the dressed gluon propagator $D(k)\propto
 \alpha_s(k^2)/k^2$ would introduce unwanted long-range Van-der-Waals
 forces between hadrons.

 \section{GRIBOV CONFINEMENT}


 Returning to the task of constructing consistent QFT dynamics of
 non-Abelian gauge fields, we must conclude that, in spite of many
 attempts, the problem of {\em Gribov copies}\/ (``Gribov horizon'',
 ``Gribov uncertainties'')
 remains essentially open today.


 By the mid-80's Gribov decided, however, to change direction and 
 not pursue {\em pure gluodynamics}. He did it not because of severe
 difficulties in describing the fundamental domain in the functional
 space: he always had his ways around technical obstacles.
 Gribov convinced himself (though not yet the physics community at
 large) that the solution to the confinement problem lies not in the
 understanding of the interaction of ``large gluon fields'' but
 instead in the understanding of how the QCD dynamics can be arranged
 as to prevent the non-Abelian fields from growing real big.
 There was a deep reason for this turn, which he formulated in the
 following words:
\begin{quote} 
  ``I found I don't know how to bind massless bosons''
\end{quote}
(read: how to dynamically construct {\em glueballs}\/).  

 As for fermions, there is a corresponding mechanism provided by the
 Fermi-Dirac statistics and the concept of the ``Dirac sea''.
 Spin-$\frac12$ particles, even massless which are difficult to
 localize, can be held together simply by the fact that, if pulled
 apart, they would correspond to the free-fermion states that are {\em
 occupied}\/ as belonging to the Dirac sea.

\subsection{Perturbative Light Quark Confinement}

 As a result of the search for a possible solution of the confinement
 puzzle Gribov formulated for himself the key ingredients of the
 problem and, correspondingly, the lines to approach it:
\begin{itemize}
\item
    The question of interest is not of ``a'' confinement, but that of
    ``the'' confinement in the real world, namely, in the world with
    two very light quarks -- $u$ and $d$ -- whose Compton wave lengths
    are much larger than the characteristic confinement scale
    ($m_q\sim 5-10\,\MeV \ll 1\,\GeV$).
\item
    No mechanism for binding massless {\em bosons}\/ (gluons) seems to
    exist in Quantum Field Theory (QFT), while the Pauli exclusion
    principle may provide means for binding together massless {\em
    fermions}\/ (light quarks).
\item
    The problem of ultraviolet regularization may be more than a
    technical trick in a QFT with apparently infrared-unstable
    dynamics: the ultraviolet and infrared regimes of the theory may
    be closely linked. Example: the pion field as a Goldsone boson
    emerging due to spontaneous chiral symmetry breaking (short
    distances) and as a quark bound state (large distances).
\item
    The Feynman diagram technique has to be reconsidered in QCD if one
    goes beyond trivial perturbative correction effects.  Feynman's
    famous $i\epsilon$ prescription was designed for (and is
    applicable only to) the theories with stable perturbative vacua.
    To understand and describe a physical process in a confining
    theory, it is necessary to take into consideration the response of
    the vacuum, which leads to essential modifications of the quark
    and gluon Green functions.\footnote{The proper technology lies in a
    generalisation of the Keldysh diagram technique designed to
    describe kinetics out of equilibrium.}
\end{itemize}
 Existence of light quarks is crucial for the Gribov confinement
 scenario.  It is clear without going into much mathematics that the
 presence of light quarks is sufficient for preventing the colour
 forces from growing real big: 
%
%
 light quarks in the vacuum are eager to screen any separating colour
 charges and turn dragged apart heavy quarks into a pair of blanched
 $D$-mesons.

 The question becomes quantitative: how strong is strong? How much of
 a tension does one need to break the vacuum and organize such a
 screening?

\subsection{Supercritical Binding}

 In a pure perturbative (non-interacting) picture, the empty fermion
 states have {\em positive energies}, while the {\em
 negative-energy}\/ states are all filled.
With account of interaction the situation may change, {\em provided}\/
two {\em positive-energy}\/ fermions (quarks) were tempted to form a
bound state with a {\em negative}\/ total energy. 
In such a case, the true vacuum of the theory would contain {\em
positive kinetic energy}\/ quarks hidden inside the {\em negative
energy}\/ pairs, thus preventing positive-energy quarks from flying
free.

 A similar physical phenomenon is known in QED under the name of
 supercritical binding in ultra-heavy nuclei.
 Dirac energy levels of an electron in an external static field
 created by the large point-like electric charge $Z>137$ become {\em
 complex}. This means instability.  Classically, the electron ``falls
 onto the centre''.  Quantum-mechanically, it also ``falls'', but into
 the Dirac sea.

 In QFT the instability develops when the energy $\epsilon$ of an
 empty atomic electron level drops, with increase of $Z$, below
 $-m_ec^2$.  An $e^+e^-$ pair pops up from the vacuum, with the vacuum
 electron occupying the level: the super-critically charged ion decays
 into an ``atom'' (the ion with the smaller positive charge, $Z-1$)
 and a real positron
$$
 A_Z \>\Longrightarrow A_{Z-1} + e^+\,, \qquad \mbox{for}\>
 Z>Z_{\mbox{\scriptsize crit.}}
$$ 
 Thus, the ion becomes unstable and gets rid of an excessive electric
 charge by emitting a positron~\cite{PM45}.
%
 In the QCD context, the increase of the running quark-gluon coupling
 at large distances replaces the large $Z$ of the QED problem.

 Gribov generalised the problem of supercritical binding in the field
 of an infinitely heavy source to the case of two massless fermions
 interacting via Coulomb-like exchange.  He found that in this case
 the supercritical phenomenon develops much earlier. Namely, a {\em
 pair of light fermions}\/ interacting in a Coulomb-like manner
 develops supercritical behaviour if the coupling hits a definite
 critical value~\cite{Lund}
\begin{equation}
\label{eq:acr}
 \frac{\alpha}{\pi} > \frac{\acr}{\pi} = 1-\sqrt{\frac{2}{3}}\,.
\end{equation}
 In QCD one has to account for the colour Casimir operator. Then the
 value of the coupling above which restructuring of the perturbative
 vacuum leads to chiral symmetry breaking and, likely, to
 confinement~(see \cite{conf1} and references therein), translates
 into
\begin{equation}
\label{eq:acrQCD}
  \frac{\acr}{\pi} = C_F^{-1}\left[\, 1-\sqrt{\frac{2}{3}}\,\right]
  \simeq 0.137\,.   \qquad C_F=\frac{N_c^2-1}{2N_c}.
\end{equation}
 This number, apart from being easy to memorize, has another important
 quality: it is numerically small. Gribov's ideas, being understood
 and pursued, offer an intriguing possibly to address all the
 diversity and complexity of the hadron world from within the field
 theory with a reasonably small effective interaction strength.

\subsection{Gribov Equation}

 Gribov constructed the equation for the quark Green function as an
 approximation to the general corresponding Schwinger-Dyson equation.
 This approximation took into account the most singular
 (logarithmically enhanced) infrared and ultraviolet effects due to
 quark-gluon interactions and resulted in a non-linear differential
 equation which possesses a rich non-perturbative structure.

 An amazing simplicity of the Gribov construction makes one wonder,
 why such an equation had not been discovered some 20 years earlier
 when a lot of effort was applied in a search for non-perturbative
 phenomena of the superconductivity type in QED (Nambu--Jona-Lasinio;
 Baker--Johnson; Fomin--Miransky et al.).

 Take the first order self-energy diagram $\Sigma_1(q)$: a fermion
 (quark/electron) with momentum $q$ virtually decays into a quark
 (electron) with momentum $q'$ and a massless vector boson
 (gluon/photon) with momentum $k=q-q'$:
\begin{equation}
\label{eq:Sigma}
  \Sigma_1(q) = [C_F]\frac{\alpha}{\pi} \int \frac{d^4q'}{4\pi^2 i}
  \left[\, \gamma_\mu \, G_0(q')\, \gamma_\mu\, \right] \> D_0(q-q')
  ,\qquad D_0(k) = \frac{1}{k^2+i\epsilon},
\end{equation}
 with $G$ and $D$ the fermion and boson propagators, respectively.
 The corresponding Feynman integral diverges linearly at
 $q'\to\infty$. To kill the ultraviolet divergences (both linear and
 logarithmic), it suffices to {\em differentiate}\/ it twice over the
 external momentum $q$.

 The first Gribov's observation was that $1/k^2$ of the boson
 propagator happens to be the Green function of the four-dimensional
 Laplace operator,
\[
   \partial_\mu^2 \frac{1}{(q-q')^2+i\epsilon} = -4\pi^2\, i\delta(q-q'),
   \qquad \partial_\mu \equiv \frac{\partial }{\partial q_\mu},
\]
 where $\partial_\mu$ denotes the momentum differentiation.
 Therefore, the operation $\partial_\mu^2$ applied to
 Eq.~\eqref{eq:Sigma} takes away the internal momentum integration and
 leads to an algebraic expression which is {\em local}\/ in the
 momentum space, $ k=q-q'=0$,
\begin{equation}
\label{eq:Born}
   \partial_\mu^2 \Sigma_1(q) \>=\> 
g\, \gamma_\mu \,
    G_0(q)\, \gamma_\mu\,, \qquad g=\left\{ 
\begin{array}{rl}
{\displaystyle \frac{\alpha}{\pi}} & \mbox{for QED}, \\
{\displaystyle C_F\frac{\alpha_s}{\pi}} & \mbox{for QCD}.
\end{array} \right.
\end{equation}
 This is the ``Born approximation''. With account of higher order
 radiative corrections, the first thing that happens is that the bare
 fermion propagator $G_0$ dresses up, $G_0(q)\to G(q)$, and so do the
 Born vertices $\gamma_\mu\to \Gamma_\mu(q,q,0)$. The second crucial
 observation was that the exact vertex function $\Gamma_\mu(q,q-k,k)$
 describing the emission of a {\em zero momentum}\/ vector boson,
 $k_\mu\equiv 0$, is not an independent entity but is related with the
 fermion propagator by the Ward identity,
\begin{equation}
   \Gamma_\mu(q,q,0) = -\partial_\mu G^{-1}(q)\,.
\end{equation}
 This statement is {\em literally}\/ true in Abelian theory (QED),
 and, after some reflection, {\em can be made}\/ true in the
 non-Abelian case (QCD) as well.

Thus, we
arrived to the Gribov equation for the quark Green
function~\cite{Lund,conf1}
\begin{equation}
\label{eq:BHeq}
  \partial_\mu^2 G^{-1}(q) = g\, \partial_\mu G^{-1}(q)\, G(q)\,
  \partial_\mu G^{-1}(q) +
\ldots \,,
\end{equation}
where $\ldots$ stand for less singular $\cO{g^2}$ integral terms.


 In principle, the whole PT series expansion may be constructed for
 the right hand side in terms of exact Green functions (and their
 momentum derivatives).  In particular, with account of the first
 subleading terms Eq.~\eqref{eq:BHeq} becomes an integro-differential
 equation and its r.h.s.\ may be represented in the following graphic
 form:
$$
 \partial_\mu^2 G^{-1}(q) \>=\>  \begin{picture}(0,0)%
\includegraphics{secord.pstex}%
\end{picture}%
\setlength{\unitlength}{3947sp}%
\begingroup\makeatletter\ifx\SetFigFont\undefined%
\gdef\SetFigFont#1#2#3#4#5{%
  \reset@font\fontsize{#1}{#2pt}%
  \fontfamily{#3}\fontseries{#4}\fontshape{#5}%
  \selectfont}%
\fi\endgroup%
\begin{picture}(4149,944)(-11,367)
\end{picture}
 + \ldots ,
$$
 with black dots standing for the momentum
 gradient of the inverse propagator -- the exact zero-momentum boson
 emission vertex~\cite{ideology}.
 Yet another set of higher order corrections makes the coupling run,
 $g\to g(q^2)$. In the $\abs{q^2}\to\infty$ limit the QCD coupling
 vanishes due to the asymptotic freedom, and Eq.~\eqref{eq:BHeq} turns
 into the free equation, $\partial_\mu^2 G^{-1}=0$, whose general
 solution has the form
\begin{equation}\label{eq:GBorn}
 G^{-1}(q) = Z_0^{-1} \left[\, (m_0- 
 \hat{q}) + \frac{\nu_1^3}{q^2} +\frac{\nu_2^4\,\hat{q}}{q^4}
 \,\right] \qquad\qquad
\left(\hat{q}\equiv \gamma_\mu q^\mu \right).
\end{equation}
 This general perturbative solution has two new arbitrary parameters
 $\nu_1$ and $\nu_2$ in addition to the familiar two (bare mass $m_0$
 and the wave function renormalization constant $Z_0$), since the
 master equation is now the {\em second}\/ order differential
 equation, unlike in the standard renormalization group (RG) approach.

 The new terms are singular at $q^2\to 0$. Therefore in QED, for
 example, we simply drop them as unwanted, thus returning to the RG
 structure. Such a prescription, however, exploits the knowledge that
 nothing dramatic happens in the infrared domain, so that the real
 electron in the physical spectrum of the theory, whose propagation we
 seek to describe, is inherently that very same object that we put
 into the Lagrangian as a fundamental bare field.

 Not so clear in an infrared unstable theory (QCD). Here we better
 keep all four terms in Eq.~\eqref{eq:GBorn}, wait and see.

 At large virtualities, $\abs{q^2}\gg m^2$, the two additional terms
 can be looked upon as power suppressed
 corrections that emerge due to non-trivial structure of the QCD
 vacuum. The corresponding dimensional parameters can be directly
 linked with the non-perturbative vacuum condensates,
$$
  \nu_1^3 \propto \lrang{\bar\psi\psi}, \qquad  \nu_2^4   \propto \lrang{\alpha_s
 F^a_{\mu\nu}F^a_{\mu\nu}},
$$
%
 which
 constituted the core of the ``ITEP sum rules'' famous for 
 successful phenomenology of
 low-lying hadron resonances, see~\cite{ITEP}.


 Moreover, in the finite momentum region, $\abs{q^2}\sim m^2$, where
 all four terms have to be treated on the same
 footing,
%
 Gribov found {\em bifurcation}\/ -- a non-perturbative solution --
 emerging in Eq.~\eqref{eq:BHeq} {\em if}\/ the coupling in the
 infrared region exceeded the critical value Eq.~\eqref{eq:acrQCD}.
 The new phase corresponds to {\em spontaneously broken chiral
 symmetry}.
 This means that given a supercritical coupling in the infrared, the
 quark Green function may possess a non-trivial mass operator even in
 the chiral limit of vanishingly small bare (ultraviolet) quark mass
 $m_0\to 0$~\cite{Lund,conf1}.

\subsection{Critical Coupling and Chiral Symmetry Breaking}

 In this section we suggest a simple algebraic exercise that should
 help a reader to see the origin of the critical coupling and to
 appreciate the Gribov equation as a tool for grasping dynamical
 breaking of the chiral symmetry.
 
 Let us introduce
\begin{equation}
\label{eq:Anudef}
  A_\nu(q)\>\equiv\> 
\partial_\nu G^{-1}(q)
\cdot G(q) .
\end{equation}
 Observing that 
$$
  \partial_\nu^2 G^{-1} - \partial_\nu\left(\partial_\nu G^{-1}\cdot
G\right) = -\partial_\nu G^{-1}\cdot \partial_\nu G =  \left(\partial_\nu G^{-1}\cdot
G\right)  \left(\partial_\nu G^{-1}\cdot G\right),
$$
 the master equation Eq.~\eqref{eq:BHeq} can be cast in the following
 compact form,
\begin{equation}
\label{eq:BH2}
 \partial_\nu A_\nu + (1-g)\, A_\nu A_\nu =0.
\end{equation}
 Using the standard representation for the fermion propagator in terms
 of the wave function renormalization factor $Z$ and the running mass $m$,
$$
  G^{-1}(q) = Z^{-1}(\xi)\left[\,m(\xi)-\hat{q} \,\right], \qquad 
  \xi=\ln q \equiv \ln \sqrt{q^2},
$$ 
and introducing 
{\em anomalous dimensions}\/ 
\begin{equation}
\label{eq:andims}
  \Gamma = \frac{\dot{Z}}{Z}, \qquad \Gamma_m = \frac{\dot{m}}{m}
  \qquad \left(\dot{f}\equiv \frac{df}{d\xi}\right),
\end{equation}
it is straightforward to derive an explicit expression for the
``vector field'' $A_\nu$: 
\begin{equation}
\label{eq:Anuex}
 A_\nu = \frac1q\left[ \sigma_{\nu\mu} n^\mu + \left(1-\Gamma\right) n_\nu -
\frac{m}{q}(\Gamma_m
-\Gamma) \hat{n}n_\nu \right]\cdot\frac{1+\hat{n}m/q}{1-(m/q)^2}.
\end{equation}
Here
$$
  n_\nu = \frac{q_\nu}{q},\>\> n^2=1; \quad
\sigma_{\nu\mu}=\frac12(\gamma_\nu\gamma_\mu-\gamma_\mu\gamma_\nu), \>\>
\sigma_{\nu\mu}\sigma_{\mu\rho}= -3g_{\nu\rho}.
$$ 
 In the region of relatively large momenta, $q\gg m$, we have
$$ 
 q\cdot A_\nu \>=\> \left[\, \sigma_{\mu\nu}n^\mu + (1-\Gamma)n_\nu\, \right]
\>+\> \frac{m}{q}\left(\gamma_\nu - \Gamma_m\hat{n}n_\nu \right) + \cO{\frac{m^2}{q^2}}. 
$$ 
Applying
this approximation to Eq.~\eqref{eq:BH2} gives
\bminiG{AAappr}\label{AA1}
q^2 \partial_\nu A_\nu &\simeq& \left[\, 2(1-\Gamma) -\dot{\Gamma}\,\right]
-\left(\dot{\Gamma}_m +\Gamma_m^2+2 \right)\frac{m}{q} \cdot \hat{n};\\
\label{AA2}
q^2  A_\nu A_\nu       &\simeq& \Bigl[\,(1-\Gamma)^2-3\,\Bigr]
\>
+ \> 2(1-\Gamma)(1-\Gamma_m) \frac{m}{q} \cdot \hat{n}.
\emini
Equating the scalar and $\hat{n}$ terms in Eq.~\eqref{eq:BH2} produces
then a system of coupled differential equations for anomalous
dimensions:
\begin{eqnarray}
\label{eq:an1}
 \dot{\Gamma} &=& 
(1-g)(1\!-\!\Gamma)^2 +2(1\!-\!\Gamma) -3(1-g), \\
\label{eq:an2}
\dot{\Gamma}_m &=&  -\left(\Gamma_m^2+2 \right) +2(1-g)(1\!-\!\Gamma)(1-\Gamma_m).
\end{eqnarray}
It is its stable point $\dot{\Gamma}= \dot{\Gamma}_m=0$ which
determines the ultraviolet anomalous dimensions $\Gamma^*$ and $\Gamma_m^*$. 
The first equation $\dot{\Gamma}=0$ is self-contained and results in
\begin{equation}\label{eq:Gamma_sol}
 (1-g)(1-\Gamma^*) = \sqrt{3(1-g)^2+1}-1, \qquad \Gamma^*=\frac{2-\sqrt{3(1-g)^2+1}-g}{1-g},
\end{equation}
where the sign of the square root was chosen such as to select among
the two fixed points in Eq.~\eqref{eq:an1} the {\em stable}\/
one.\footnote{The second solution corresponds to
$\Gamma^{*}(g\!=\!0)=4$ and describes renormalization of the power
suppressed $Z^{-1}\nu_2^4/q^4$ term of the free fermion propagator
Eq.~\eqref{eq:GBorn}.}
 The anomalous dimension of the mass in Eq.~\eqref{eq:an2} is driven
 by that of the wave function, since from $\dot{\Gamma}_m=0$ we have
\begin{equation}\label{eq:Gammam_exp} 
 \Gamma_{m\pm}^* = -(1-g)(1-\Gamma) \pm \sqrt{[(1-g)(1\!-\!\Gamma) + 1]^2-3}\,.  
\end{equation} 
Substituting $\Gamma$ from Eq.~\eqref{eq:Gamma_sol}, for the stable
point we finally obtain
\begin{equation}\label{eq:Gammam_sol}
 \Gamma_m^*= \Gamma_{m+}^* \>=\> 1-\sqrt{3(1-g)^2+1} + \sqrt{3(1-g)^2-2}. 
\end{equation}
Expressions Eqs.~\eqref{eq:Gamma_sol}--\eqref{eq:Gammam_sol}
are {\em non-perturbative}\/ since they have emerged from 
{\em non-linear}\/ 
Eq.~\eqref{eq:BHeq}.
The first term of the series expansion in $g$,
$$
 \Gamma^*(g)= \frac12\,g + \cO{g^2}, \qquad \Gamma_m^*(g) = -\frac32\,g + \cO{g^2},
$$
coincides with the known one-loop perturbative expression for the
anomalous dimension of the (Feynman gauge) fermion wave function and
of the (gauge invariant) mass operator, respectively.


%
 Now we can readily see 
 how the {\em critical coupling}\/ emerges that have been announced
 above in Eq.~\eqref{eq:acr}. Indeed, when $g$ reaches the value
 $g_{\crit}= 1- \sqrt{\frac23}$, the running 
 mass becomes {\em complex}\/ due to the term $\sqrt{3(1-g)^2-2}$ in
 Eq.~\eqref{eq:Gammam_sol}.

 This signals instability. In fact, at this point the two anomalous
 dimensions of Eq.~\eqref{eq:Gammam_exp} become equally important,
 $\Re \Gamma_{m\pm}^*(g=g_{\crit})= -(\sqrt{3}-1)$.  The mass operator
 $m(\xi)$ remains real but ceases to be monotonic and starts to
 oscillate with $\xi=\ln q$. It is this oscillatory behaviour that
 allows one to construct a symmetry breaking solution corresponding to
 $m_0=0$. Its mass operator is regular at $q=0$ and decays fast in the
 ultraviolet, $m(\xi)\propto \exp(-2 \xi)\propto 1/q^2$, where the
 chiral symmetry gets dynamically restored. 

 For explicit construction of this chiral symmetry breaking solution,
 its properties and physics of accompanying Goldstone states
 see~\cite{conf1}.


\subsection{Quarks, Pions and Confinement}

 As far as {\em confinement}\/ is concerned, the approximation
 Eq.~\eqref{eq:BHeq} turned out to be insufficient. A numerical study
 of the Gribov equation carried out by C.~Ewerz showed~\cite{Carlo}
 that the corresponding quark Green function does not possess an
 analytic structure that would correspond to a confined object.

 Given the dynamical chiral symmetry breaking, however, the Goldstone
 phenomenon takes place bringing pions to life. In his last
 paper~\cite{conf2} Gribov argued that the effects that Goldstone
 pions induce, in turn, on the propagation of quarks is likely to lead
 to confinement of light quarks and, as a result, to confinement of
 any colour states.

 The approximate equation for the Green function of a massless quark,
 which accommodates a feed-back from Goldstone pions reads~\cite{conf2}
\begin{eqnarray}
\label{eq:BHmod}
 \partial_\mu^2 G^{-1}(q) &=&\> g(q)\, \partial_\mu G^{-1}(q)\, G(q)\,
  \partial_\mu G^{-1}(q) \nonumber\\
 && - \frac{3}{16\pi^2 f_\pi^2}
\left\{i\gamma_5,G^{-1}(q)\right\}\,G(q)\, \left\{ i\gamma_5,G^{-1}(q)\right\}.
\end{eqnarray}

 It is important to notice that since pions have emerged {\em
 dynamically}\/ in the theory, their coupling to quarks is not
 arbitrary but is tightly linked with the quark propagator itself
 (search for an anti-commutator of $\gamma_5$ with $G^{-1}$ in
 Eq.~\eqref{eq:BHmod}).  Moreover, the pion--axial current transition
 constant $f_\pi$ is not arbitrary either, but has to satisfy a
 definite relation which, once again, is driven by the behaviour of
 the exact quark Green function:
\begin{eqnarray}
\label{eq:fpi}
f_\pi^2 &=&\> \frac18 \int\frac{d^4q}{(2\pi)^4i} \Tr\left[\>
\left\{i\gamma_5,G^{-1}\right\}\,G\, \left\{ i\gamma_5,G^{-1}\right\} G
\left(\partial_\mu G^{-1}\,G\right)^2\>\right] \nonumber \\
&&+ \>\>\frac1{64\pi^2f_\pi^2} \int \frac{d^4q}{(2\pi)^4i} \Tr\left[\> 
\left(\left\{i\gamma_5,G^{-1}\right\}\,G\right)^4\>\right].
\end{eqnarray}

 The second of the two papers~\cite{conf1,conf2} concluding Gribov's
 study of the light-quark supercritical confinement theory remained
 unfinished.  It ends abruptly in the middle of the discussion of the
 most intriguing question, namely, what is the meaning, and practical
 realization, of unitarity in a confining theory.

 The modified Gribov equation Eq.~\eqref{eq:BHmod} still awaits a
 detailed study aiming at the analytic structure of its solutions.

\subsection{Gluon Sector}

 Another important open problem is to construct and to analyse an
 equation for the gluon similar to that for the quark Green function,
 from which a consistent picture of the coupling $g(q)$ rising above
 the critical value in the infrared momentum region should emerge.


 The difficulty with the gluon sector of the theory lies in separating
 the running coupling effects from an unphysical gauge dependent phase
 that are both present in the gluon Green function.
 To analyse renormalization of the gluon Gribov has used
 in~\cite{conf1} the Duffin--Kemmer formalism (also known as ``linear
 formalism'') which treats the gluon potential $A_\mu^a$ and the field
 strength $F_{\mu\nu}^a$ as independent variables. Renormalization
 properties of the gauge invariant correlator $\lrang{FF}$ gives then
 a direct access to the running coupling. The Duffin--Kemmer technique
 being plagued with artificial divergencies, this attempt did not
 result however in an equation for $\alpha_s$ in a closed form.

 The solution may lie in using the ``background gauge'' or in
 exploiting the unitarity motivated ``pinch'' 
 technique developed by J.~Papavassiliou, N.J.~Watson and others
 (see~\cite{unitar} and references therein), as in both
 approaches the QCD coupling is directly related with the
 renormalization of the gluon Green function.

\section{QED AT SHORT DISTANCES}

 At the same time, to construct a smart
 equation for the running coupling
 poses no problem in an Abelian gauge theory where the boson
 propagator is gauge invariant.

 Gribov has carried out this programme for QED. 
 Here the coupling becomes supercritical at extremely short
 distances. Gribov's analysis aimed at resolving the long-standing
 problem of the ``Landau pole'' in the running QED coupling.  The
 ``Landau ghost'' in the photon propagator at academically large
 momenta
 seems to be a formal problem so long as QED is actually a part of a
 broader field theoretical scheme.
 Nevertheless, its resolution is extremely important since it
 demonstrates what type of new, non-perturbative, phenomena in QFT one
 might expect when the strength of the coupling becomes large
 (dynamical symmetry breaking and appearance of Goldstone states,
 condensation, confinement, etc.).

 Gribov did not finish the paper entitled ``Quantum Electrodynamics at
 Short Distances''. Handwritten notes he left behind, deciphered and
 translated into English by J.~Nyiri, appeared in~\cite{RBook_QED}.
 It is important to mention that no attempt has been made by the
 editorial team to ``debug'' these notes.\footnote{This was done on
 purpose.  By correcting omnipresent mistakes/misprints in obvious
 places the editors could have misled a potential reader into putting
 too much trust into not-so-obvious derivations and conclusions. As
 the matter stands now, it should be not {\em read}\/ but rather {\em
 worked through}\/ with a pencil and an A4-pad at hand.
 We strongly encourage to do so everyone 
 willing to play a $\frac13$--of--a--Rosetta--Stone quest.}

 Here we will reproduce essential steps of the part of Gribov's
 analysis that directly addressed the ``Moscow zero'' problem. 
 By so doing we hope to be able to ignite enough curiosity so that our
 readers will be tempted to look themselves into more delicate issues
 (such as appearance of light super-bound states, condensation, the
 Goldstone phenomenon) that were also considered in~\cite{RBook_QED}.

\subsection{Equation for Photon Polarization Operator}

 Electron loop Feynman diagrams for the polarization operator
 $\Pi_{\mu\nu}(k)$ for a photon with momentum $k$ diverges
 quadratically. Therefore in order to obtain a convergent integral
 (and thus a finite answer) we need to differentiate it {\em thrice}\/
 over $k$. At the same time, conservation of current dictates
\begin{equation}\label{eq:PIdef}
 \Pi_{\mu\nu}(k) = \left(k_\mu k_\nu-g_{\mu\nu}k^2\right)\cdot \Pi(\xi), \qquad \xi=\ln k,
\end{equation}
where the factor $\Pi$ diverges logarithmically in the ultraviolet; in
the Born approximation,
$$ 
  \Pi_0 = e_0^2\cdot \mbox{const}\cdot
  \ln\frac{\Lambda_{\mbox{\scriptsize UV}}}{k}.  
$$
 Let us apply successively two differential momentum operations to Eq.~\eqref{eq:PIdef}:
\begin{equation}\label{eq:PIdiff}
 \partial_\mu\equiv \frac{\partial}{\partial k_\mu}, \quad 
 \partial_\mu\Bigl[\partial_\nu \Pi_{\mu\nu}(k)\Bigr] =
 \partial_\mu \Bigl[\, 3k_\mu \Pi(\xi)\,\Bigr] \>=\> 3(4\Pi\,+\dot{\Pi})
\end{equation}
(recall that the dot marks $\xi$--derivative).  To get rid of the
remaining logarithmic divergence it suffices to differentiate
Eq.~\eqref{eq:PIdiff} over $\xi$.  This is equivalent to applying the
operator $k_\sigma\partial_\sigma$ which produces 
$$ 
\dot{\Pi}+ \frac14\ddot{\Pi} \>=\> \frac{(k_\sigma\partial_\sigma)}{12}\cdot
\, \partial_\mu \partial_\nu\, \Pi_{\mu\nu}(k).  
$$ 
 The l.h.s.\ can be expressed directly through
 the running QED coupling. Indeed, observing that, by definition, 
$$
 g(\xi)= \frac{\alpha(\xi)}{\pi}= \frac{e^2(\xi)}{4\pi^2} =
\frac{e_0^2}{4\pi^2}\cdot\frac{1}{1+\Pi(\xi)}, 
$$ 
we may write
\begin{equation}\label{eq:newRG}
\left(\frac{d}{d\xi}+ \frac14 \frac{d^2}{d\xi^2}\right) \frac{1}{g}
\>=\>  \frac{(k_\sigma\partial_\sigma)}{6}\cdot \frac{2\pi^2}{e_0^2}\>  
 \partial_\mu \partial_\nu\, \Pi_{\mu\nu}(k).
\end{equation}
 As for the r.h.s.\ of Eq.~\eqref{eq:newRG}, we treat $\Pi_{\mu\nu}$
 as a sum of Feynman diagrams and apply the first two differentiations
 {\em diagrammatically}. This results in
%
\begin{equation}\label{eq:Loop_deriv}
\frac1{e_0^2} \, \partial_\mu \partial_\nu\, \Pi_{\mu\nu}(k) \> = \> \sum\quad
\begin{picture}(0,0)%
\includegraphics{floop.pstex}%
\end{picture}%
\setlength{\unitlength}{3947sp}%
\begingroup\makeatletter\ifx\SetFigFont\undefined%
\gdef\SetFigFont#1#2#3#4#5{%
  \reset@font\fontsize{#1}{#2pt}%
  \fontfamily{#3}\fontseries{#4}\fontshape{#5}%
  \selectfont}%
\fi\endgroup%
\begin{picture}(2709,1651)(1069,-1179)
\end{picture}

\end{equation}
where crosses mark momentum derivative of the bare electron
propagator,
\begin{equation}\label{eq:G_deriv}
 \partial_\mu G_0(q+k) =  \partial_\mu \frac1{m_0-(\hat{q}+\hat{k})} =
 G_0 \gamma_\mu  G_0; \quad {\bf\times} = - \partial_\mu G_0^{-1}= \gamma_\mu.
\end{equation}
 (Integral over the loop momentum $q$ diverges only logarithmically
 which justifies integration by parts used to derive Eq.~\eqref{eq:Loop_deriv}.)
\def\Bigdot{\begin{picture}(0,0)%
\includegraphics{dot.pstex}%
\end{picture}%
\setlength{\unitlength}{3947sp}%
\begingroup\makeatletter\ifx\SetFigFont\undefined%
\gdef\SetFigFont#1#2#3#4#5{%
  \reset@font\fontsize{#1}{#2pt}%
  \fontfamily{#3}\fontseries{#4}\fontshape{#5}%
  \selectfont}%
\fi\endgroup%
\begin{picture}(104,259)(2334,40)
\end{picture}
}
 After differentiation, the overlapping divergencies disappear that
 plagued the original Schwinger--Dyson equation for the polarization
 operator. Therefore, higher order corrections combine to {\em
 renormalize}\/ four fermion propagators, $G_0(q)\to G(q)$, both
 external vertices, $\gamma_\mu\to\Gamma(q,q+k,k)$, as well as the
 crosses:
\begin{equation}\label{eq:cross_point}
 {\bf\times}= \gamma_\mu= -\frac{\partial}{\partial q_\mu} G_0^{-1}(q)
\quad\to\quad  -\frac{\partial}{\partial q_\mu} G^{-1}(q) =
\Gamma_\mu(q,q,0) \>\equiv\> \Bigdot.
\end{equation}
We thus arrive at the renormalized graph 
\begin{equation}\label{eq:Loop_deriv1}
\frac1{e_0^2} \, \partial_\mu \partial_\nu\, \Pi_{\mu\nu}(k) \> = \> 
\begin{picture}(0,0)%
\includegraphics{floop1.pstex}%
\end{picture}%
\setlength{\unitlength}{3947sp}%
\begingroup\makeatletter\ifx\SetFigFont\undefined%
\gdef\SetFigFont#1#2#3#4#5{%
  \reset@font\fontsize{#1}{#2pt}%
  \fontfamily{#3}\fontseries{#4}\fontshape{#5}%
  \selectfont}%
\fi\endgroup%
\begin{picture}(2416,1625)(1205,-1183)
\end{picture}
 \> +\> \Delta.
\end{equation}
For the correction term $\Delta=\cO{g}$ a perturbative series
expansion can be constructed in terms of two-particle irreducible
multi-loop graphs built of exact propagators and interaction vertices,
\begin{equation}\label{eq:Loop_deriv2}
\Delta \>=\>
\begin{picture}(0,0)%
\includegraphics{floop2.pstex}%
\end{picture}%
\setlength{\unitlength}{3947sp}%
\begingroup\makeatletter\ifx\SetFigFont\undefined%
\gdef\SetFigFont#1#2#3#4#5{%
  \reset@font\fontsize{#1}{#2pt}%
  \fontfamily{#3}\fontseries{#4}\fontshape{#5}%
  \selectfont}%
\fi\endgroup%
\begin{picture}(4997,1625)(1701,-1183)
\end{picture}
 + \cO{g^2}.
\end{equation}
Feynman integral corresponding to the leading contribution in
Eq.~\eqref{eq:Loop_deriv1} reads
\begin{equation}\label{eq:loop_int}
\int\frac{d^4q}{(2\pi)^4i}
\Tr\Bigl[\,\Gamma_\mu(q,q\!+\!k,k) G(q) A_\mu(q) \Gamma_\nu(q\!+\!k,q,-k)
G(q\!+\!k) A_\nu(q\!+\!k) \,\Bigr],
\end{equation}
where we have combined the adjacent dot-operator $\Bigdot=\partial_\mu
G^{-1}(q)$ and the Green function $G(q)$ 
into $A_\mu$ of Eq.~\eqref{eq:Anudef}.

 We are interested solely in the logarithmically divergent
 contribution to Eq.~\eqref{eq:loop_int} since finite pieces, after
 applying the last differentiation $(k_\sigma\partial_\sigma)$ in
 Eq.~\eqref{eq:newRG}, will produce but negligible corrections that
 are power suppressed for large external photon momenta, $k\gg m$.
 Bearing this in mind, in the relevant integration region $k\ll
 q\ll\LUV$ we can omit $k$ in the loop propagators, $G(q+k)\simeq
 G(q)$, and in the exact photon vertices,
$\Gamma_\mu(q,q\!+\!k,k) \simeq \Gamma_\mu(q,q,0) = -\partial_\mu G^{-1}(q)$,

In the large momentum region we invoke Eq.~\eqref{AA2} to obtain
$$
 \Tr\Bigl[\, A_\mu(q) A_\mu(q)\, A_\nu(q) A_\nu(q) \,\Bigr] \simeq
\frac4{q^4}\cdot \left([1-\Gamma]^2-3 \,\right)^2.
$$
Transforming the integration phase space,
$$
 \int\frac{d^4q}{(2\pi)^4i\,q^4} = \frac{1}{8\pi^2} \int d\xi, 
$$
we derive
$$
\frac{2\pi^2}{e_0^2} \partial_\mu  \partial_\nu \Pi_{\mu\nu}(k)
\>\simeq\>  \int_{\ln k}^{\ln \LUV} d\xi\, \frac{\Tr [\quad ]}{4}.
$$
Substituting 
into Eq.~\eqref{eq:newRG} and taking the $\xi$-derivative we finally
arrive at
\begin{equation} \label{eq:RGalpha}
 \left(\frac{d}{d\xi}+ \frac14 \frac{d^2}{d\xi^2}\right) \frac{1}{g}
 \>\simeq\> -\frac{1}{6} \left([1-\Gamma]^2-3\right)^2.
\end{equation}
 This second order differential equation generalises the standard RG
 equation,
$$
  \frac{d}{d\xi}  {g}^{-1} \>=\> \beta(g), 
$$
with the $\beta$-function expressed via anomalous dimension of the
(Feynman gauge) electron wave function, $\Gamma=\Gamma^*(g)$.

\subsection{Small and Large Coupling Regimes}

In the small coupling regime we have $\Gamma=\cO{g}\ll 1$ and
Eq.~\eqref{eq:RGalpha} gives
$$
 g^{-1} \>\simeq g_0^{-1} - \frac{2}{3}\,\ln k,
$$
which is the standard one-loop expression for the running QED coupling
that formally develops ``Landau pole'' in the deep ultraviolet, $k\sim
m_e\exp(\frac{3\pi}{2}\cdot 137)$. However, for $g=\cO{1}$ there is no
reason to neglect $\Gamma$.
The latter is given by
\begin{equation} 
  \Gamma(g) \>=\> 1- \frac{3(1-g)}{1+\sqrt{1+3(1-g)^2}} .
\end{equation}
From this representation, which is equivalent to the original
Eq.~\eqref{eq:Gamma_sol}, we observe that $\Gamma$ is monotonically
increasing with $g$, crosses unity at $g=1$ and tends to
%
$$ 
 \Gamma(g) \>\stackrel{g\to\infty}{=} \> (1+\sqrt{3}) - g^{-1} + \cO{g^{-2}} .
$$
We observe that in the large coupling limit the r.h.s.\ of
Eq.~\eqref{eq:RGalpha} becomes small,
$$
 \left(\frac{d}{d\xi}+ \frac14 \frac{d^2}{d\xi^2}\right) \frac{1}{g}
 \>\simeq\> -\frac{1}{6}\cdot \frac{(2\sqrt{3})^2}{g} =   - \frac{2}{g},
$$
resulting in {\em logarithmic increase}\/ of the coupling at large
photon virtualities:
\begin{equation} \label{nonpert}
  g(\xi) \>\simeq\> 2\cdot \xi =  \ln \abs{k^2} \gg 1.
\end{equation}
Landau singularity has disappeared. Gribov explained this phenomenon
as a compensation between the contributions to vacuum polarization
from {\em magnetic moment}\/ of electron [$(\sigma_{\mu\nu})^2=-3$] by
that of its {\em charge}\/ [$(1-\Gamma)^2$] in the non-perturbative
regime of large anomalous dimensions.

Numerically large coupling $g\gg 1$ does not mean that the interaction
becomes really {\em strong}. If this were the case, the
approximation based on neglecting higher order terms $\Delta$ in
Eq.~\eqref{eq:Loop_deriv1} that was adopted in Eq.~\eqref{eq:RGalpha}
would have been undermined.
Gribov argued that since adding a photon is accompanied by additional
factors $A_\mu$ in each vertex, the relative magnitude of radiative
corrections, in spite of $g\gg 1$, may turn out to be finite, $g\cdot
(A_\mu)^2=\cO{1}$.


 In fact, there is no need to analyse the structure of $\Delta$ for
 $g\gg 1$, which regime is actually of little interest. Indeed, as
 have we discussed above, already at $g=\cO{1}$ a new phenomenon
 occurs, which is supercritical binding of fermion pairs. It leads to
 appearance of scalar/pseudoscalar ``mesons'' and changes drastically
 the behaviour of the Abelian theory in the ultraviolet.

\subsection{Composite Higgs as a Super-critically Bound $\bar{t}t$ Pair}

 As an interesting byproduct of his supercritical quark confinement in
 the {\em infrared}, Gribov considered in 1994 an application of the
 supercritical picture to the weak sector of the Standard Model in the
 {\em ultraviolet}.

 It is widely believed that the couplings of the basic interactions
 $U(1)_{\mbox{\scriptsize Y}}$, $SU(2)_{\mbox{\scriptsize weak}}$ and
 $SU(3)_{\mbox{\scriptsize strong}}$ merge at some Grand Unification
 scale, above which the underlying dynamics becomes essentially
 different.
 However, if there is a ``Super-Great Desert'' instead, the Abelian
 hyper-charge interaction constant $g_Y$ keeps growing with scale
 leading to spontaneous symmetry breaking, non-zero vacuum expectation
 value
 $V$, generation of $W/Z$ masses
 and appearance of Higgs boson as a super-bound state of top quarks.
 All these phenomena being inter-connected, this scenario allows one
 to relate $m_H$ and $m_t$ with the symmetry breaking parameter
 $V\simeq 246 \GeV$, the Weinberg angle and the values of the
 couplings $\alpha_s$ and $\alpha_{e.m.}$ at the top quark mass
 scale~\cite{TopHiggs}.

 The mass of such composite Higgs particle (whose binding is similar
 to that of a deuteron in the zero-radius limit of nuclear forces)
 would have been $m_H=(170\pm 7)\,\GeV$, given the experimental top
 quark mass $m_t\simeq 175\GeV$ (and varying $\alpha_s$ between $2m_t$
 and $\frac12 m_t$).

\section{CONCLUSIONS}

  Hadron phenomenology has accumulated a very impressive dossier of
  puzzles and hints, ranging from unexplained regularities in hadron
  spectroscopy to soft ``forceless'' hadroproduction in hard
  processes.

 The reason why one keeps talking, 30 years later, about {\em puzzles
 \&\ hints}, about {\em constructing}\/ QCD rather than routinely {\em
 applying}\/ it, lies in the conceptually new problem one faces when
 dealing with a non-Abelian theory with unbroken symmetry. We have to
 learn to master a Quantum Field Theory whose dynamics is
 intrinsically unstable in the infrared domain so that the objects
 belonging to the physical spectrum of the theory (supposedly,
 colourless hadrons, in the QCD context) have no direct one-to-one
 correspondence with the fundamental fields the microscopic Lagrangian
 of the theory is made of (coloured quarks and gluons).

 In these circumstances we don't even know how to formulate at the
 level of the microscopic fields the fundamental properties of the
 theory, such as conservation of probability (unitarity) and
 analyticity (causality). Indeed,
\begin{itemize}
\item
  What does {\em\bf Unitarity}\/ imply for confined objects?
\item
  How does {\em\bf Causality}\/ restrict quark and gluon Green
  functions and their interaction amplitudes?
\item
  What is the {\em\bf Mass}\/ of an INFO -- [well] Identified [but]
  Non-Flying Object?~\footnote{The issue of quark masses is
  particularly damaging since a mismatch between quark and hadron
  thresholds (whatever the former means) significantly affects
  predicting the yield of heavy-flavoured hadrons in hadron
  collisions, markedly at LHC.}
\end{itemize}

 Understanding the confinement of colour remains an open issue.
 The very problem can be formulated in various terms, ranging from a
 $10^6$\$\ worth a rigorous mathematical proof of the existence of a
 mass gap in pure gluodynamics all the way to developing down-to-earth
 but theory motivated practical recipes for cooking the spectrum of
 hadrons and predicting their production properties.

 Among many 
a contribution
 to the problematics of non-Abelian gauge theories -- proper
 quantization of Yang--Mills fields, the origin of asymptotic freedom,
 the nature of instantons and physics of quantum anomalies, etc., --
 the key ingredient of the Gribov conception of Quantum Chromodynamics
 was 
 setting up 
 the problem of the confinement of colour as that of {\em light
 quarks}.

\vspace{0.5 cm}

 Gribov works on gauge theories and, in particular, all his papers,
 talks and lectures devoted to anomalies and the QCD confinement
 (including the lectures that were translated into English for the
 first time)
 were collected and recently published\footnote{
 The book can be ordered at http://www.prospero.hu/gribov.html}
 in~{\em Gauge Theories and Quark Confinement}~\cite{RBook}.


 From these papers, an interested reader will be able to follow the
 derivation of the Gribov equation and to study the properties of its
 (perturbative and non-perturbative) solutions, as well as to
 formulate and pursue the open problems awaiting analysis and
 resolution.  Pedagogical lectures he gave in 1992 in
 Orsay~\cite{Orsay} will give an opportunity to grasp the physical
 picture of the supercritical binding which includes an anti-intuitive
 notion of an ``inversely populated'' Dirac sea and to think about
 phenomenological aspects of the light quark confinement scenario.

  Needless to say, the Gribov approach and recent developments
  reviewed here did not yet provide a consistent recipe for dealing
  with QCD at large distances. Nevertheless, an impressive progress
  has been made, and Gribov's ideas continue to inspire many of those
  who have contributed to it.

%
%

\end{document}